# Record cryogenic cooling in ferroelectric hafnia proximity induced via Mott transition

**Authors:** Jalaja M A[1]*, Shubham Kumar Parate[1], Binoy Krishna De[1], Sai Dutt K[1], Pavan Nukala[1]*

**Affiliations:**
[1] *Centre for Nano Science and Engineering, Indian Institute of Science, Bengaluru, 560012, India*

*Corresponding author. Email: jalaja.gvc@gmail.com (J.M.A), pnukala@iisc.ac.in (P.N.)

**Abstract:** On-chip refrigeration at cryogenic temperatures is becoming an important requirement in the context of quantum technologies and nanoelectronics. Ferroic materials with enhanced electrocaloric effects at phase transitions are good material candidates for the same. By exploiting the Mott metal-insulator transition (MIT) of $TiO_x(N_y)$, the bottom electrode, we engineer a depolarization field controlled reversible polar to non-polar phase transition in thick La-doped hafnia (40 nm). This transition occurs between ~125 and 140 K and produces giant negative pyroelectric and electrocaloric effects. Refrigeration metrics were estimated between 120 to 200 K, with a peak refrigerant capacity of 25 kJ $Kg^{-1}$ (2 kJ $Kg^{-1}$), peak isothermal entropy ΔS~ 8 kJ $Kg^{-1}$ $K^{-1}$ (0.5 kJ $Kg^{-1}$ $K^{-1}$) and adiabatic $\Delta T_{cooling}$ ~ 106 K (11 K) at ~140 K and 5 MV $cm^{-1}$ (0.5 MV $cm^{-1}$, and these are the largest reported in any electrocaloric system. Our work fundamentally proposes design guidelines to induce significant solid-state refrigeration through proximity effects, even at cryogenic temperatures relevant to quantum technologies.

**One-sentence Summary:**

Negative electrocaloric effect-based refrigeration at cryogenic temperatures in wake-up-free hafnia film induced via reversible Mott-transition in the electrode.

## Introduction

On-chip refrigeration technology addresses heat dissipation-related issues to sustain advancements in nanoelectronics. Cryogenic on-chip cooling, in particular, can reduce the footprint of refrigeration and will be very crucial for quantum technologies. Magneto, electro and elastocaloric effects in ferroic materials can be exploited for such applications. Thermodynamically, a change in order parameter (OP) with temperature and a change in entropy (S) with an external stimulus are converse effects. In other words, if OP is polarization (magnetization), and the external stimulus is an electric field (magnetic field), then pyroelectric (pyromagnetic) coefficients ($\pi$) are the same in magnitude as electro (magneto) caloric effects as shown in equation 1.

$$\pi = \left(\frac{dP}{dT}\right)_E = \left(\frac{\partial S}{\partial E}\right)_T \quad (1)$$

Here P is the spontaneous polarization, T is the temperature and E is the electric field.

Ferroic materials can produce giant caloric effects in the vicinity of phase transitions. Metrics such as isothermal entropy change ΔS, adiabatic cooling ($\Delta T$) whose differential forms are given by equations 2, 3 and refrigeration capacity (RC) (equation 4) qualify the efficacy of refrigeration when run through an ideal thermodynamic cycle.

$$dT = -\frac{T}{C(E,T)\rho}\left(\frac{\partial P(E,T)}{\partial T}\right) dE \tag{2}$$

$$dS = \frac{dP(E,T)}{dT} dE \tag{3}$$

$$RC = \int_{T_1}^{T_2} \Delta S \, dT \tag{4}$$

Here, $\rho$ is the mass density; and $C(E,T)$ is the heat capacity as a function of field and temperature.

Solid-state cooling technologies began with advancements in magnetocaloric materials. Magnetic materials such as $Fe_{49}Rh_{51}$ ($\Delta T \sim 9.2$ K, RC ~ 125 J $kg^{-1}$)(*1*) and $Gd_5Si_2Ge_2$ ($\Delta T \sim 15$ K, RC ~255 J $kg^{-1}$)(*2*), show a magnetocaloric effect generally at and above room temperature. In the $RVO_4$ (R=Rare earth) series, $GdVO_4$ was found to be a promising candidate for cryogenic refrigeration, capable of producing a substantial temperature reduction from 20.3 K to 3.5 K upon demagnetizing from 50 kOe (*3*).

Electrocaloric (EC) materials, particularly ferroic perovskites and polymers are gaining attention recently. $\Delta T$ values up to 45.3 K and RC values as large as 2125 J $Kg^{-1}$ at room temperature are reported on $Pb_{80}Ba_{20}ZrO_3$ perovskite thin-film (*4*). Multilayer capacitor (MLC) devices made of $PbSc_{0.5}Ta_{0.5}O_3$ (PST)(*5*) and lead-free BTO(*6*, *7*) produce efficient EC cooling with $\Delta T$ values of approximately 7.1 K and 13 K and RC values 320 J $Kg^{-1}$and 70 J $Kg^{-1}$, respectively at and above room temperatures. At a device level, asymmetric ferroelectric tunnel junctions (FTJs) based on $BaTiO_3$ as the tunnel barrier also were shown to achieve large significant figures of merit such as $\Delta T$~4.6 K and RC~6 J $Kg^{-1}$(*8*) at room temperature. In recent breakthroughs at a system level, prototype pyroelectric harvesting modules and electrocaloric regenerator with optimal structural design and cooling of 13 K were demonstrated on the PST MLC systems.

Large negative electrocaloric effect (ECE)s was reported in certain antiferroelectrics and relaxors, where the application of an electric field increases the entropy ($\left(\frac{\partial P}{\partial T}\right) = \left(\frac{\partial S}{\partial E}\right) > 0$ ) of the system. Thin film heterostructures based on $PbZr_{0.53}Ti_{0.47}O_3/CoFe_2O_4$ produce a large EC cooling of 52 K and RC~ 4888 J $Kg^{-1}$ (*9*) at 182 K. Negative electrocalorics show adiabatic cooling upon application of an electric field, whereas positive electrocalorics cool upon removal of the electric field. Exploring both negative and positive (dual) EC systems aims to maximize cooling efficiency(*10*, *11*).

In the context of refrigeration at cryogenic temperatures, incipient ferroelectrics such as $SrTiO_3$ (STO) and $KTaO_3$ were explored with $\Delta T$~0.3 K, between 4–15 K (*12*, *13*). Strain-engineered STO films grown on $DyScO_3$ show a second-order phase transition at 243 K, resulting in large electrocaloric effects compared to bulk STO with a peak Delta S of ~0.35 J $Kg^{-1}K^{-1}$ *(14).*

Integrating conventional perovskite-based ferroelectrics into modern CMOS technology faces challenges in material compatibility, scalability, and toxicity. CMOS-compatible, unconventionally ferroic (ferroelectric and antiferroelectric) hafnia and zirconia-based systems present an opportunity for the exploration of on-chip coolers integrable into nanoelectronic circuits. $Hf_{0.2}Zr_{0.8}O_2$ shows large positive ECEs with $\Delta T$ =13.4 K and RC > 488 J $Kg^{-1}$ (*15*). Significant refrigeration metrics at room temperature were recently demonstrated in Si-doped hafnia systems with pyroelectric coefficients of –1300 $\mu C\ m^{-2}\ K^{-1}$, $\Delta T$~9.5 K, and RC value of 203 J $Kg^{-1}$ (*16*). Antiferroelectric $ZrO_2$ films of 8 nm thickness showed a very large negative ECE with $\Delta T$= -31 K, RC > 501 J $Kg^{-1}$ at an ambient temperature of 413 K (*17*). Dual EC behaviour was shown in Al-doped $HfO_2$ thin films as a function of concentration of Al; the negative ECE ($\Delta T$= -7.4 K and RC ~ 57 J $kg^{-1}$) with a low (5.6 mol%) concentration of Al was observed at 80 K and the positive ECE ($\Delta T$= 5.7 K and RC ~ 32 J $kg^{-1}$) with a high (6.9 mol%) concentration of Al was observed at 295 K (*18*). It may be noted that significant ECEs were reported in hafnia-based systems, even without taking advantage of any temperature-dependent phase transitions (ferro to para, for e.g.). Refrigeration applications using ferroic hafnia/zirconia also will require demonstrating significant ECEs on samples with larger active volumes, prompting the usage of La or Y as dopants, that can stabilize ferroic nature even in thicker samples (*19-21*).

In this study, we couple a robust thick ferroelectric La doped Hafnia (LHO) layer to a Mott insulator, a bottom electrode (BE), that exhibits reversible Mott insulator to metal transition (IMT) at 140 K. By doing so, we proximity induce reversible non-polar to polar phase transitions in the LHO layer, and engineer a record negative EC response: $\Delta S$~ 8 kJ $Kg^{-1}\ K^{-1}$ (0.5 kJ $Kg^{-1}\ K^{-1}$), RC~ 25 kJ $Kg^{-1}$ (2 kJ $Kg^{-1}$), and $\Delta T$ ~ -106 K (11 K) corresponding to the applied field 5 MV $cm^{-1}$ (0.5 MV $cm^{-1}$) respectively, at the cryogenic transition temperatures. Our LHO thick films are prepared via a scalable and cost-effective solution processing technique and show robust ferroelectricity ($P_r$= 29 $\mu C/cm^2$, cyclability >$10^9$ cycles) at room temperature.

**Results and Discussions**

TiN (170 nm) was deposited using reactive magnetron sputtering on Si (100). On this template, 40 nm to 60 nm thick 7 % LHO was synthesized using chemical solution deposition and subsequently crystallized at 650 °C (methods). Thinner LHO layers (<20 nm) remained amorphous even after 650 °C annealing (Fig. S1). Film uniformity was confirmed from cross-sectional FESEM and STEM (Fig. S2) measurements, with a surface roughness of 1.5 nm (Fig. S3). Grazing incidence XRD shown in Fig. 1a, shows a (111) reflection at 2θ=30.42 ° corresponding to an orthorhombic (o-) phase in LHO. Notably, any Bragg reflections corresponding to the non-polar monoclinic (m) phase (for instance at 2θ =28.52 °) are conspicuously absent. Peaks at 2θ = 36.11 °, 41.29 °, 62.82 ° match well with those of $Ti_4O_7$ (magneli phase, ICDD card no.77–1392), and are completely different from the cubic phase reflections of as-sputtered TiN (Fig. 1a) (*22*). Thus our post-annealing step at 650 °C oxidized the as-deposited TiN layer into a conducting $TiO_x(N_y)$ layer resembling the metallic I-phase of $Ti_4O_7$, ($I\bar{1}$) phase (Unit cell volume, V=465.25 Å$^3$ at room temperature)(*23*).

Cross-sectional STEM analysis confirms that LHO crystallizes in a single o-phase with various domains (Fig.1(b, c)). To gain more insights into the chemistry and symmetry of the orthorhombic phase of LHO and the bottom electrode, spectroscopic tools including EDS, Raman, XAS and XPS were performed. Quantitative EDS analysis performed over various regions in LHO layer confirms a spatially uniform cationic composition of 7% La in Hf (Fig. S2(c)). EDS chemical mapping confirms that the BE is $TiO_x(N_y)$, with an almost negligible presence of Nitrogen (Fig. S2(d)). Raman spectra can clearly distinguish between the various polymorphs of hafnia, especially the structurally similar ones, and hence are a good fingerprint for precise phase determination(*24*, *25*). Fig. 1d shows the room temperature Raman spectrum from our films and the corresponding mode of deconvolution. Raman mode at 471 $cm^{-1}$ is the signature of the polar o-phase ($Pca2_1$) of LHO (*24*) and other signature peaks at 144, 234, and 610 $cm^{-1}$ correspond to various modes of $Ti_4O_7$ (*26*, *27)*. The origin of various Raman peaks and their identification are presented in Table S1.

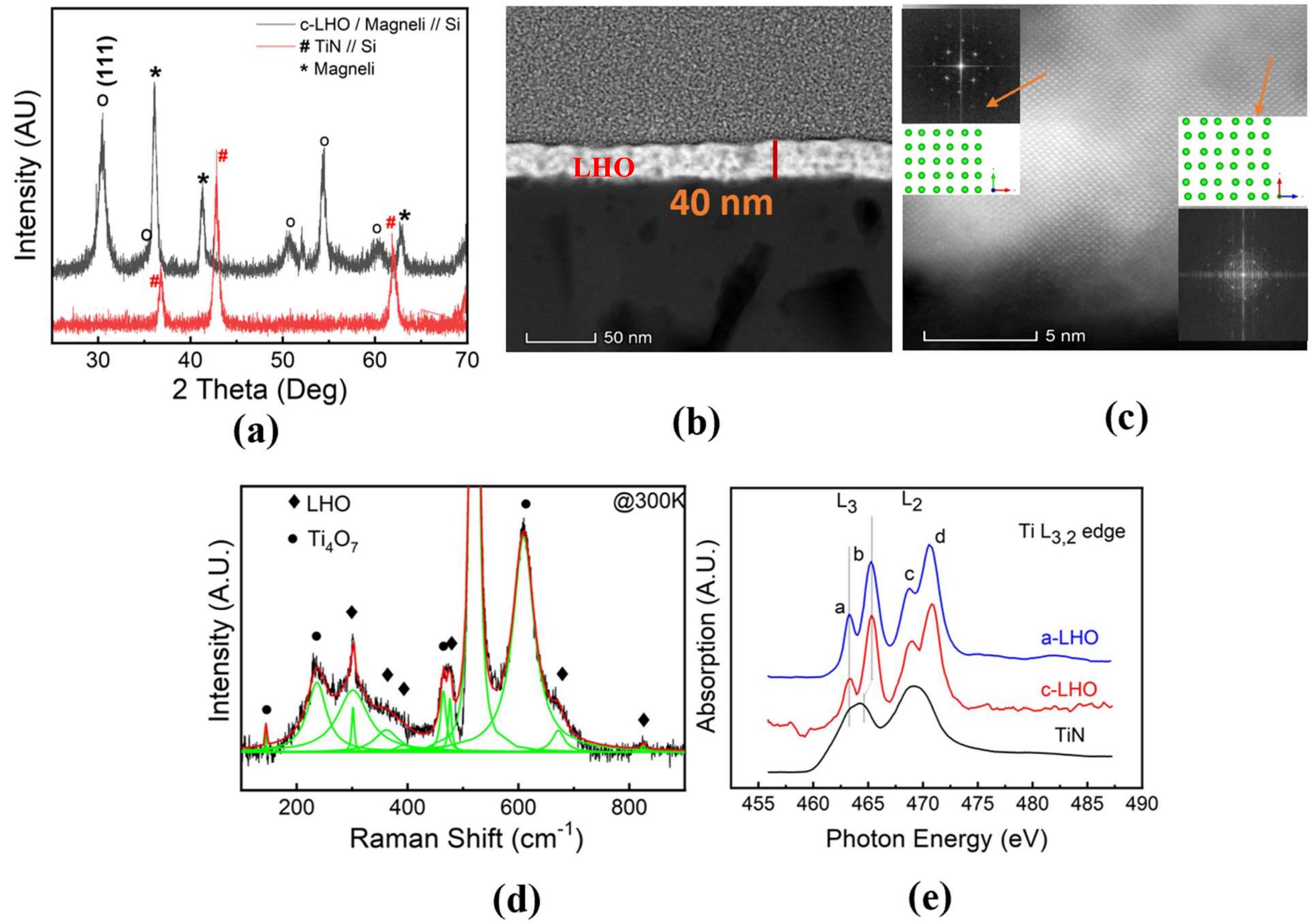

**Fig.1. Structure, microstructure and composition of LHO/magneli//Si stacks at room temperature**

**(a)** Room temperature GIXRD patterns of reference TiN//Si samples (red) and LHO/Magneli//Si heterostructures. **(b)** STEM cross-sectional image of the TiN (top electrode)/LHO/Magneli stack indicating the thickness of the LHO layer as 40 nm. **(c)** HAADF-STEM cross-sectional image showing orthorhombic domains in LHO, with zone axis: [010] on the left (LD) and [001] on the right (RD). (Inset) shows corresponding FFT and schematic of real space cationic atomic arrangement pattern of both LD (left) and RD (right domain). **(d)** Raman spectrum of various modes in LHO/Magneli layers at room temperature, and corresponding peak fits and **(e)** XAS spectra comparing Ti $L_{2,3}$ edges of amorphous (a-) LHO/magneli//Si (blue), crystalline (c-) LHO/magneli //Si (red) and reference TiN//Si layers (black).

The oxidation state of Ti in the bottom electrode was probed through XAS. Fig. 1(e) shows comparative $L_3$ and $L_2$ edges of Ti XAS spectra collected from a reference TiN layer, the bottom electrode (BE) in amorphous LHO (a-LHO) (<20 nm)/BE//Si and crystalline LHO (c-LHO) (40 nm)/BE//Si samples (see methods). The $L_3$ ($L_2$) edges are further split into two more finer features, a and b (c and d) edges (Fig. 1e). These features orginate due to the electronic transitions from Ti 2p core level to the crystal field split 3d bands i.e. $t_{2g}$ and $e_g$ bands, respectively. The BE of both crystalline and amorphous LHO/$TiO_x(N_y)$//Si show a clear $e_g$-$t_{2g}$ band split (ΔE=Eb-Ea) of ~2.04 eV(*28*). From the ΔE calibration presented in Fig. S4, we conclude that the Ti in this layer exists as a mixture of $Ti^{3+}$ and $Ti^{4+}$ ions (Fig. S4), also consistent with a magneli phase. Note that pure TiN does not show any clear $e_g$-$t_{2g}$ split. Additionally, depth-resolved XPS spectra collected from the BE of our crystalline LHO samples, indicate the presence of a small amount of Nitrogen (Fig. S5). From all these analyses, we conclude that while LHO crystallizes in the polar $Pca2_1$ phase, the electrode resembles the magneli phases ($Ti_4O_7$ structurally) mixed with some nitrogen. From here on we will refer to the BE as a magneli layer.

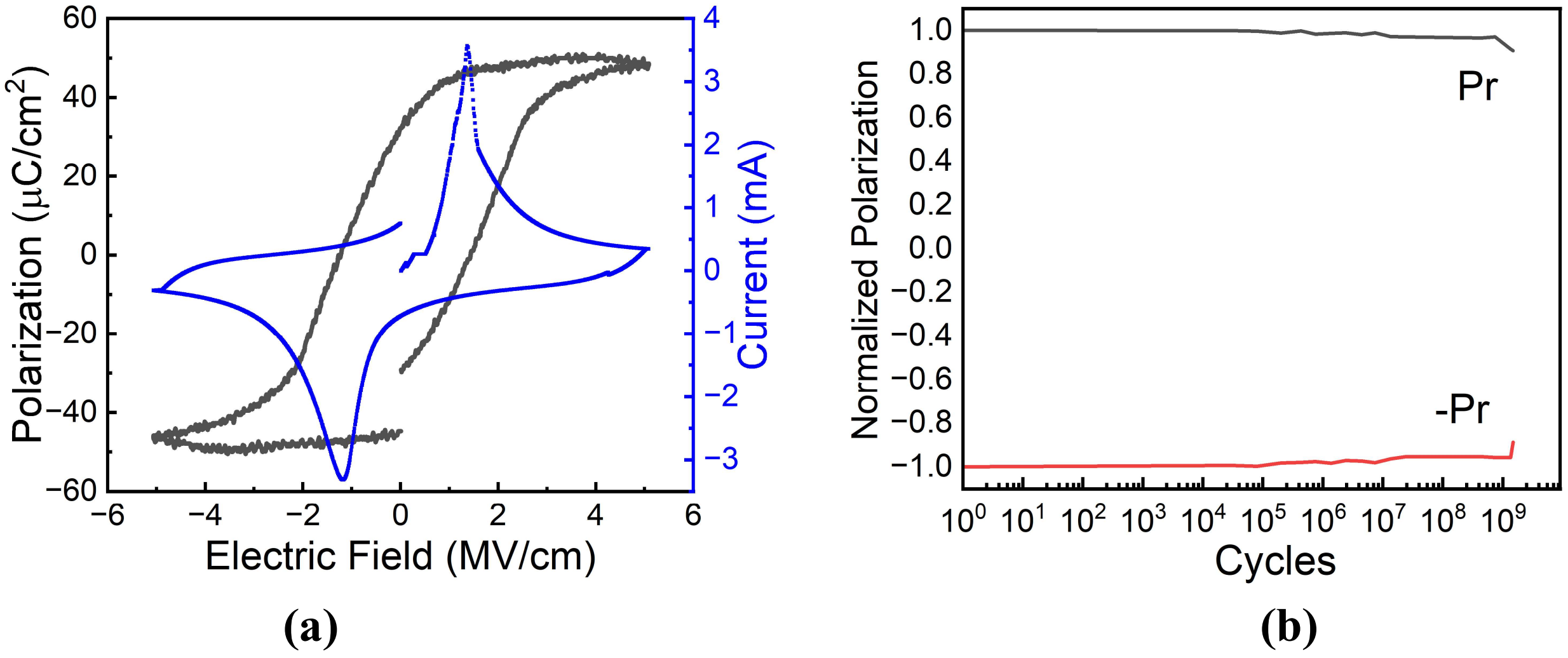

**Fig.2. Ferroelectric characteristics of TiN/c-LHO/Magneli capacitors at room temperature** **(a)** Room temperature P-E hysteresis loop and corresponding ferroelectric switching current for capacitors tested at 3 kHz, **(b)** Endurance testing at 1 MHz, showing normalized remanant polarization with cycles for the TiN/LHO/magneli devices.

By patterning TiN as the top electrode, metal-insulator-metal (TiN/LHO/Magneli) capacitors were fabricated. Ferroelectric switching with a large $P_r$ ~29 μC $cm^{-2}$ (at 297K) was measured on these capacitors (Fig. 2a). While robust ferroelectricity in doped hafnia is typically observed in thinner samples (<13 nm) (*29-31*), dopants such as Y, La can stabilize ferroelectric phase (in a phase mixture) in thicker samples (>20nm) (*19-21*)

Furthermore, our devices are wake-up-free and can be cycled up to $>10^9$ cycles, beyond which effects of fatigue set in (Fig. S6). Wake-up is a commonly observed effect in hafnia-based ferroelectrics (*19, 30*). There are a few reports on wake-up-free hafnia on thinner films, most consistently on PLD-based epitaxial hafnia, but these are all limited to low thicknesses (<13 nm) *(30-32)*. Recent works have shown wake-up-free ferroelectricity in chemical solution-deposited thick LHO films, albeit with $P_r$ ~ 25 μC $cm^{-2}$, attributed to uniform and homogenous distribution of oxygen vacancies throughout the film (*19*, *20*).

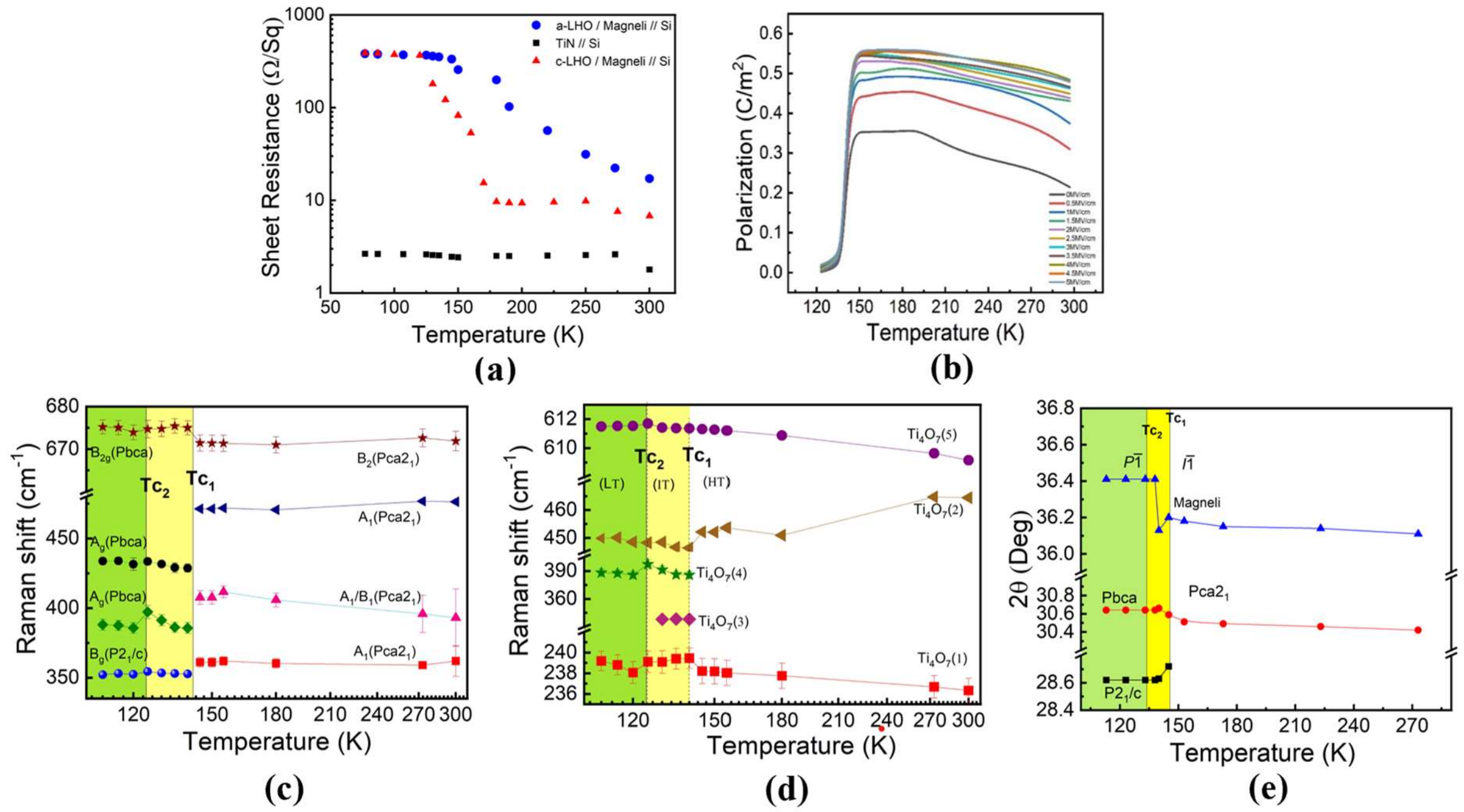

**Fig.3. Temperature evolution of ferroelectric, transport and structural properties of various layers through the metal-insulator transition of the magneli layer**
**(a)** Comparison of variation of the sheet resistance of the bottom electrode as a function of temperature for reference TiN//Si, magneli layer in a-LHO/magneli//Si and c-LHO/magneli//Si films measured during heating. **(b)** Temperature-dependent remanant polarization values at different applied fields. Temperature evolution of Raman peaks for **(c)** Crystalline LHO and **(d)** magneli phase in the c-LHO/magneli//Si sample. $Tc_1$ and $Tc_2$ are the transition temperatures. **(e)** variation of o-LHO, m-LHO and magneli peaks with varying temperatures.

Next, we measured the sheet resistance of the magneli layer as a function of temperature in a van der Pauw geometry (Fig.3a). While as-deposited TiN (on intrinsic Si) displays a metallic conductivity from 77 K to 300 K, the magneli phase interfaced with crystalline LHO shows a sharp reduction in sheet resistance by more than two orders of magnitude between 125 and 175 K (heating data is shown in Fig S7). This is a reversible first-order Mott IMT, similar to the one reported in $Ti_4O_7$ (*33*, *34*) (with $T_c$=149 K). When the magneli phase is interfaced with amorphous LHO, the IMT starts at a slightly larger temperature (145 K), suggesting tunability of $T_c$, with its exact value possibly depending on the precise $Ti^{3+}/Ti^{4+}$ ratio *(35)* and the content of nitrogen in the magneli phase. It may be noted that for T in the range 77 to 125 K, even though the magneli layer is in an insulating phase, it still behaves as an electrode, albeit with more contact resistance (ρ~1 μΩ m, similar to a bad metallic electrode such as Indium Tin Oxide).

To understand the effect of IMT on the ferroelectric properties of our capacitors, we performed temperature-dependent P-E loop measurements (Fig. S8 (a,b)), capturing both heating and cooling cycles from 77 K to 300 K (Fig.S8(c)). Above 143 K, we see a gradual reduction in $P_r$, as is standard for conventional ferroelectrics. Interestingly, no ferroelectric PE loops were observed below 135 K (Fig. S8(b)). This is concomitant with the transformation of metallic magneli phase into an insulating magneli phase. Furthermore, temperature hysteresis in $P_r$ upon heating and cooling about the phase transition is narrow (Fig. S8 (c)). Polarization (electrical displacement) at various electric fields shows a steep increase with temperature from 125-140 K (negative pyroelectric), followed by a gradual decrease up to room temperature (positive pyroelectric) (Fig.3(b)). The pyroelectric coefficient ($\pi$=-$\frac{dP}{dT}$) is plotted as a function of temperature in Fig S9, and its absolute value at 140 K is larger than 0.12 C $m^{-2}$ $K^{-1}$ at various fields (Fig. S9(a)). These are one of the largest reported values of π for any pyroelectric material near phase transitions (*9)*. Fig S9(b) shows a bar chart comparing the values of |π| reported for various standard pyroelectric materials, and ferroic materials at phase transitions.

The voltage-dependent capacitance (C-V) characteristics (Fig. S10 (a)) show butterfly-shaped non-linear characteristics consistent with ferroelectric switching beyond 140 K, with zero field capacitance increasing with temperature. Below 135 K, however, the ferroelectric butterflies disappear, and C-V curves show dielectric behaviour. The temperature vs capacitance plots (Fig. S10(b)) also shed light on similar behaviour with a drop in capacitance values beyond 143 K.

The disappearance of ferroelectricity in the capacitors at the IMT of the magneli phase, also results in corresponding structural transformations in both layers. Temperature-dependent *in situ* Raman spectroscopy and GIXRD from 77 K to 300 K were performed on our films to probe these transformations. The temperature evolution of various Raman modes of magneli and LHO layers are shown in Fig. 3(c) and 3(d), respectively (see Table S1). In the range between 300 K to 140 K, (Fig. S11), only modes corresponding to polar o-phase ($Pca2_1$) of LHO ($A_1$ at 362, 393, 476 $cm^{-1}$ $A_2$ at 301 $cm^{-1}$, $B_1$ at 393 $cm^{-1}$ and $B_2$ at 671, 825 $cm^{-1}$) exist (*24*, *25*, *36,37*), and show a very weak temperature dependence. Below 140 K, the $B_2$ mode of polar o-hafnia ($Pca2_1$ at 671 $cm^{-1}$), for example, exhibits a sudden shift to a higher wavenumber at 675 $cm^{-1}$, corresponding to the $B_g$ symmetry mode of the antipolar (Pbca) orthorhombic phase. Other modes corresponding to non-polar o-phase (Pbca, $A_g$ at 388, 433 $cm^{-1}$) and non-polar monoclinic phase ($P2_1/c$) ($B_g$ at 349 $cm^{-1}$) also appear below 140 K ($Tc_1$), signifying a phase transition from polar o-phase to a mixture of non-polar o and m-phases. Careful inspection of the temperature evolution of the modes corresponding to the non-polar phases below 140 K suggests the existence of another critical temperature ($Tc_2$) at 125 K. Furthermore, the evolution of the mode of the magneli phase (identified with respect to $Ti_4O_7$, Fig. 3e), is also consistent with two phase transitions in this layer (*33*, *34*): i.e. metal to a semiconductor at $Tc_1$ and semiconductor to insulator at $Tc_2$. The 338 $cm^{-1}$ that only exists between $Tc_1$ and $Tc_2$ uniquely corresponds to a semiconducting phase of $Ti_4O_7$ (*25*).

Temperature-dependent GIXRD shows that the d(111) of the polar o-phase of LHO shifts marginally increases with temperature from 2.9272 A° at ~150 K (2θ=30.5192°) to 2.9357 A° at room temperature 2θ=30.4234°), capturing thermal expansion (Fig. 3(e), Fig. S12, Fig. S13). $d_{(111)}$ suddenly decreases from 2.9198 A° (2θ= 30.5915°) to 2.9153 A° (2θ=30.6410°) between 145 and 125 K signifying a polar o-phase ($Pca2_1$) to non-polar o-phase (Pbca) transition in response to IMT of the proximity layer. Furthermore, a non-polar monoclinic (111) peak appears at ~28.7204° ($d_{111}$=3.1060 A°) below 145 K, which further reduces to 28.6229° ($d_{111}$=3.1167 A°) below 125 K (Fig.12(a)). Also, note that Mott IMT in the magneli phase also is a first-order structural transition that results in a sudden increase of $d_{(120)}$ value from 2.465 A° (36.42°) at 125 K to 2.479 A° (36.20°) at 145 K (Fig.11(b)). Careful inspection of GIXRD data also reveals the presence of three regimes in both LHO layer and magneli layer (*33*, *34*), consistent with Raman data: (i) <125 K ($Tc_2$) where LHO is a mixture of m-phase and non-polar o-phases, and magneli phase is insulating ($I\bar{1}$) (ii) >145 K ($Tc_1$) where LHO is in a polar o-phase, and magneli layer is metallic ($I\bar{1}$) and (iii) 125-145 K, where magneli is semiconducting ($P\bar{1}$) and LHO possibly contains a mixture of $Pca2_1$, Pbca phases, in addition to a compressed monoclinic phase ($P2_1/c$, 2θ=28.62°, Fig. S12(a)). Our structural data is also very consistent with changes in sheet resistance of the magneli layer, and polarization data of the capacitors.

Proximity-induced phase transitions in LHO layer can be explained from the framework of depolarizing fields. The metallic magneli layer sustains the polarization of the LHO layer by providing sufficient screening charges to screen the depolarizing field. However, when this layer transforms into an insulator, more depolarization sets in the LHO layer. Interestingly, our results show that LHO responds to the depolarization field by reversibly transitioning into non-polar phases (o and m phases), unlike conventional ferroelectrics that form domains. Furthermore, the IMT in the magneli layer is a volume-reducing phase transition (from 467.20 $Å^3$ at 115 K to 465.64 $Å^3$ at 140 K) and thus will impose compressive stress on the LHO layer. Such compressive stresses also aid in the concomitant high-volume non-polar m-phase to the low-volume polar o-phase transition. The compressive stress on the m-phase of LHO induced by the IMT in the magneli layer is also evident from the reduction of its $d_{111}$ from 3.1167 A° when the magneli phase is insulating to 3.1060 A° at 145 K, when the magneli layer is in the semiconducting phase (see Fig 3e).

Let us recall that the pyroelectric coefficient (π) in the range between 125 and 140 K is a large negative number, whose magnitude $|\pi| > 0.12$ C $m^{-2}$ $K^{-1}$ (see table S2 for comparison with other materials). From equation 1, this also means that our system exhibits a giant converse property, i.e. negative ECE ($\frac{dS}{dE}$). Hence the refrigeration metrics *viz* adiabatic cooling (ΔT (T,E)) with application of field, isothermal entropy change (ΔS (T,E)), and RC (T,E), are also large, and we estimated them carefully starting from equations 2 and 3 (differential forms). The estimation procedure and the assumptions used are described in supplementary information (SI). We used an analytical non-linear function to fit P as a function of T. The exact form of this function changes the final estimation of ΔT (T,E). With that caveat, in Fig. 4a we present the minimum estimates of ΔT (T,E), with P (T,E) empirically fit as non-linear Weibull Cumulative Distribution functions of T at every E. We observe peak ΔT(T,E) of -11 K and -106 K at E= 0.5 and 5 MV $cm^{-1}$ respectively (T~140 K). Compared to other ferroic hafnia systems and electrocalorics at room temperature, these values are giant (Table S2), attributable to huge π (see Fig S9(b)), and the operability of our devices at fields as large as 5 MV $cm^{-1}$ (BTO based tunnel junctions, for instance, operate at 100s of kV/cm)(*8*).

ΔS (T,E) at 140 K are 8 and 0.5 kJ $Kg^{-1}$ $K^{-1}$ at E= 5 and 0.5 MV $cm^{-1}$ respectively (Fig. 4(b)). These are also record-high values. The next best ΔS reported for an EC system at room temperature is < 0.1 kJ $Kg^{-1}$ $K^{-1}$, orders of magnitude less than our devices (*38–40*). RC, a measure of heat that gets transferred from the cold end to the hot end in an ideal refrigeration cycle, is plotted as a function of T in Fig. 4c, and also displays giant values at cryogenic temperatures (>1200 J $K^{-1}$) (*9*,*17*, *38*).

Large negative ECE was also reported in $PbZr_{0.53}Ti_{0.47}O_3/CoFe_2O_4$ (PZT/CFO) heterostructures demonstrated with ΔT= -52.3 K, RC= 4888 J $Kg^{-1}$, argued to be arising out of the magnetoelectric coupling in the system that can be controlled via the arrangement of the ferroic layers (*9*). Our system shows much larger effects at cryogenic temperatures of ~125-145 K, which also seems to be tunable with the precise stoichiometry of the magneli layer.

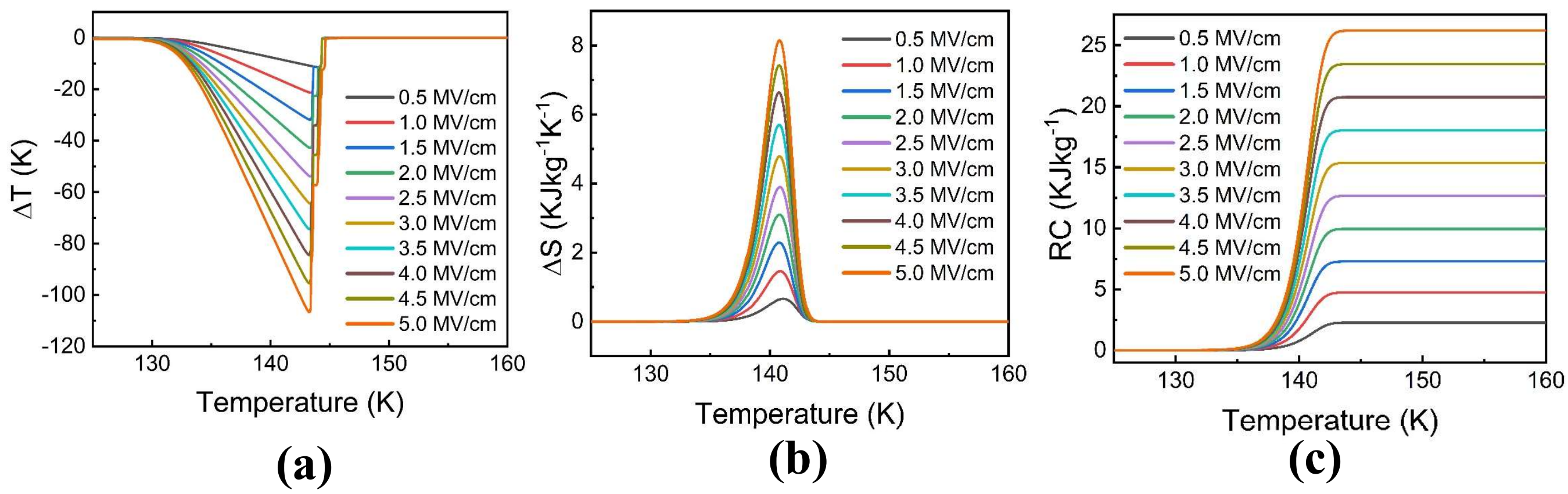

**Fig. 4. Evaluation of refrigeration metrics of our TiN/c-LHO/magneli capacitors**

Temperature-dependent **(a)** ΔT, **(b)** ΔS and **(c)** RC values with temperature under different electric fields for the TiN/LHO/magneli devices

In summary, through a metal-(semiconductor)-insulator transition in the proximity magneli layer, we engineered polar to non-polar phase transitions in ferroic LHO, via reversibly turning off and on the depolarization fields and changing the elastic boundary conditions. In turn, this induces giant pyroelectric and ECEs at these cryogenic temperatures. In addition, the ability of LHO films to operate under large fields (5 MV $cm^{-1}$), renders record refrigeration metrics for our samples.

It is important to note that hafnia-based ferroelectrics have already made an incredible mark in the microelectronics industry. However, their utility for their associated properties such as piezoelectric, pyroelectric and ECEs are found wanting, and exploration is still in its infancy. We believe that proximity-induced phase transitions and corresponding large entropy changes provide an innovative way to engineer and enhance these properties. Our approach can be generalized to obtain ECEs in a range of cryogenic temperatures, by a choice of a suitable choice stoichiometry and strain-engineered Mott insulators as the back electrode, taking us a step closer towards building on-chip refrigeration systems at ultralow temperatures.

# Supplementary Materials

## Preparation of thin films

La-doped hafnia (LHO) films were prepared by a cost-effective wet chemical method involving the preparation of the precursor sol followed by spin coating and heat treatments. Hf(IV)-2,4-pentanedionate, (Alfa Aesar, 97% purity) and Lanthanum (III) acetylacetonate hydrate (Alfa Aesar, 97% purity) were used as the ingredients for preparing the precursor sol. The cationic La % in the precursor was selected as 7%. The precursor sol containing both La and Hf cations was then spin-coated on TiN/Si bottom contact with an RPM of 4000 and pyrolyzed at 350$^o$C, and the same procedure was repeated until a desired number of layers (n) was obtained. Finally, the films were annealed at 680$^o$C. At n=12, 40 nm crystalline LHO was synthesized, and at n=3 <20 nm amorphous LHO was synthesized.

## Characterization

*XRD, XPS, XAS and Raman spectroscopy:*

X-ray diffraction: The structural and microstructural analysis of LHO thin films was carried out using x-ray diffractometry. The x-ray diffraction patterns were obtained from the Rigaku Smartlab diffractometer with $CuK_{\alpha 1}$ radiation (λ= 0.15405929 nm), parallel beam optics and point detector (scintillation counter) in grazing incidence mode with an incidence angle of 0.5 $^o$. The reference patterns for orthorhombic and monoclinic $HfO_2$ were taken from the powder diffraction files with the numbers, numbered (04-005-5597) and (00-034-0104) respectively.

X-ray photoelectron spectroscopy (XPS) measurement was performed using XPS Instrument Kratos Ultra DLD with the monochromatic Al Kα radiation (140 watts) as the x-ray source. To obtain thickness-dependent information, especially on the bottom electrode, the samples were subjected to Ar etching (depth-resolved XPS). The instrument was calibrated using Ag $3d_{5/2}$ peak of the standard Ag sample and adventitious carbon was measured at 284.6 eV.

X-ray absorption near-edge structure (XANES) experiments were performed at Ti L edge in polarized-light soft x-ray absorption beam line (BL-01) at the bending magnet source of Indus-2 synchrotron source at RRCAT, India, with total electron yield mode.

Raman spectra were obtained using a Raman spectrometer (HORIBA JOBIN VYON make, model LabRAM HR (UV) system) with a 532 nm laser excitation source. Temperature-dependent measurements were carried out using a helium cryostat in the temperature range 77 K–350 K. The temperature was monitored by a Lake Shore temperature controller. During the data acquisition, a pressure of $10^{-6}$ m bar was maintained in the cryostat.

*TEM and EDX:*

Structural analysis was done using High Angle Annular Dark Field Scanning Transmission Electron Microscope (HAADF-STEM) on Thermofisher TITAN Themis operated at 300 kV, with a 24 mrad convergence angle. Compositional analysis was performed in the same instrument equipped with quad SuperX EDS detectors.

*Film thickness measurements:*

The thickness of the films is measured using FESEM cross-sectional analysis, thickness profilometer (after making a step) and TEM cross-sectional imaging.

*Electrical measurements:*

To measure the electrical properties, TiN top electrode was sputter coated on the LHO films using a shadow mask with circular patterns of diameter 250 microns and spacing 250 microns. All the electrical measurements were performed in an out-of-plane mode with a MIM (TiN/LHO/magneli) device structure. Capacitance-voltage (C-V), and temperature-capacitance measurements were carried out using an Agilent Device Analyzer (B1500A) in a low-temperature probe station (Lake Shore Cryotronics) with a cryogenic refrigerator. C-V tests were performed with 3 kHz AC probing frequency and 50 mV amplitude. The sheet resistance of both the bottom electrode and film was measured using the van der Pauw method in the temperature range of 77 K-400 K, during cooling and heating. The ferroelectric I-V and P-E characteristics were obtained using Radiant Technologies, Inc. from 77 K-400 K at 3 kHz. The polarization switching fatigue behaviour was measured by cycling the device by applying electrical pulses, with $E_{max}$=5 MV $cm^{-1}$, and a frequency of 1 MHz. Data was read using a 3 kHz triangular pulse intermittently.

*Electrocaloric figures of merit calculations:*

The P-E data were then used to estimate the ECE. 9.680 g $cm^{-3}$ was used as the mass density (r) of hafnia *(16)* to calculate ΔT and ΔS. Specific heat capacity values as a function of temperature were obtained from the data on a tetragonal phase of hafnia (extracted from ref.*(41)*). We note that $C_E$ values and their temperature trends remain more or less the same in other phases also, justifying our assumption to take the variation of $C_E(T)$ in the t-phase as representative data for all the phases.

Generally, EC cooling is estimated from the following equation, which assumes that ΔT is very small compared to the operating temperatures.

$$\Delta T = -\int_{E1}^{E2} \frac{T}{C(E,T)\rho}\left(\frac{\partial P(E,T)}{\partial T}\right) dE$$

However, that assumption fails in our case when ΔTs are large. The following equation is a differential form and a more fundamental form of estimating ΔT.

$$\left(\frac{dT}{dE}\right) = -\frac{T}{C(E,T)\rho}\left(\frac{dP(E,T)}{dT}\right)$$

However, this is an implicit equation which can be solved numerically, given that ($\frac{dP}{dT}$) is both a function of T and E, and $C_E$ is a function of T (dependence of C on E is neglected). Non-linear curve fitting of P(T) was necessary to ensure continuity and non-diverging values of ($\frac{dP}{dT}$) and S-shaped/sigmoid functions were ideal approximations for our P(T) data at all the fields, given the sharp nature of the phase transition. Dielectric displacement was plotted as a function of temperature at different electric fields to fit for various sigmoidal functions – Langevin, Logistic, Hill, and so on. Weibull distribution functions give good overall fits of P as a function of T ($P_i(T)$) at all the different electric fields $E_i$=(0.5, 1, 1.5, 2, 2.5, 3, 3.5, 4, 4.5 and 5 MV cm$^{-1}$) and, therefore, were used to fit and generate data points. In this process, the field dependence of P, $(\frac{dP}{dT})$ was considered only at discrete points spaced at 0.5 MV cm$^{-1}$, with an assumption that P(T, $E_i$)=Pi(T) all over the interval ($E_i$-0.5, $E_i$) MV cm$^{-1}$. $\Delta T_i$ calculations were performed in intervals of 0.5 MV cm$^{-1}$ and $\Delta T(E_j)$ is estimated as $\sum_{i=1}^{j}(\Delta T_i)$. ΔS was estimated from the following differential equation.

$$dS = \frac{dP(E,T)}{dT}dE$$

Here again, $(\frac{dP}{dT})$ is described by a set of temperature-dependent non-linear functions in field intervals of 0.5 MV cm$^{-1}$. ΔS calculations were performed in intervals of 0.5 MV cm$^{-1}$ and are summed up to obtain the value of ΔS at all the electric fields. ΔS was integrated over various temperature ranges to give refrigeration capacity.

$$\Delta S_i = \int_{E_i}^{E_{i+1}} \frac{dP_i}{dT}dE$$

$$\Delta S(E_i) = \sum \Delta S_i$$

**Supplementary Figures**

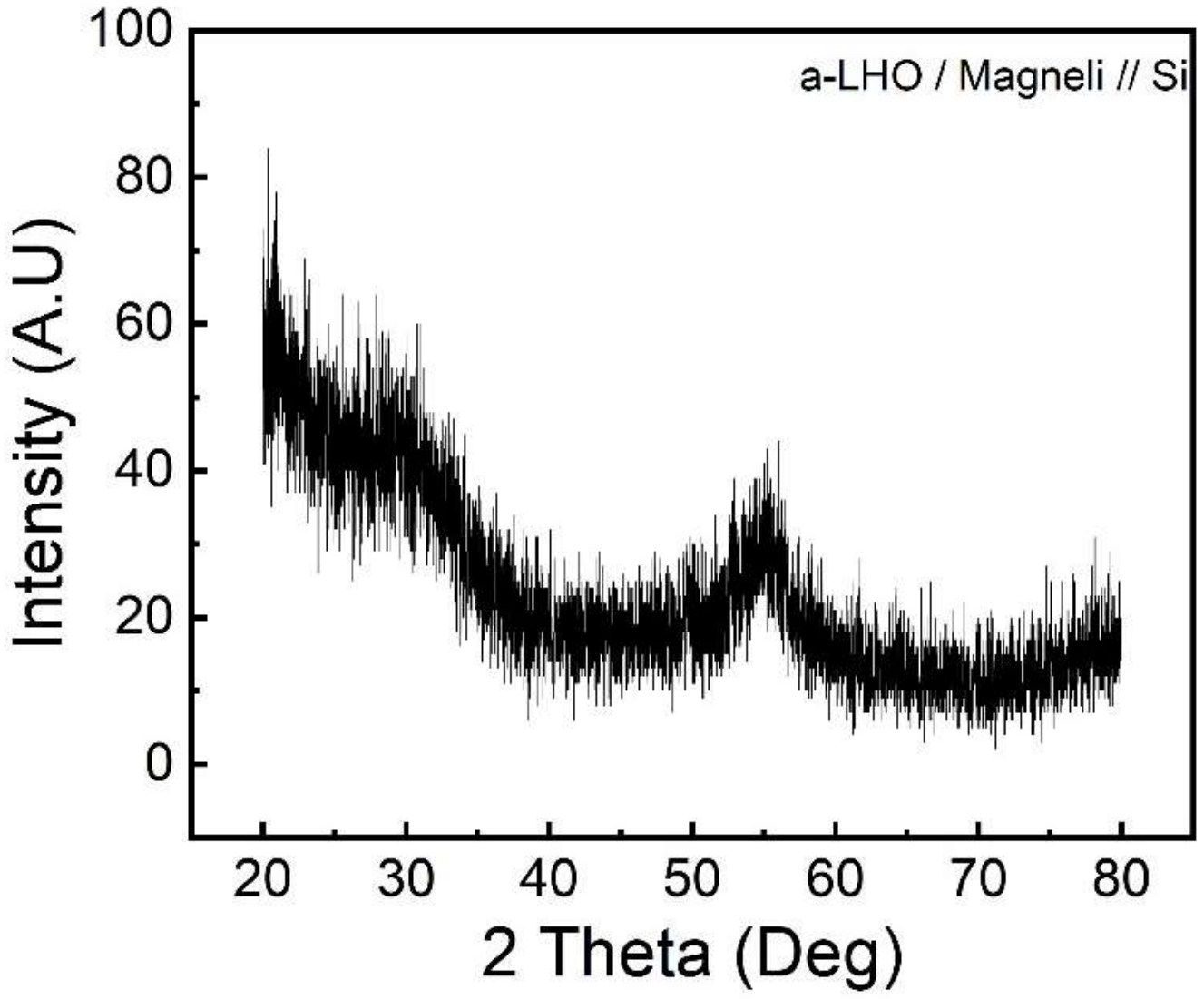

>

**Fig. S1. Amorphous LHO film characterization**

GIXRD pattern of LHO film, 20 nm thick, obtained by three layers of spin-coated and pyrolyzed precursor, followed by annealing at 680°C. These films show amorphous characteristics and are denoted as a-LHO in the manuscript.

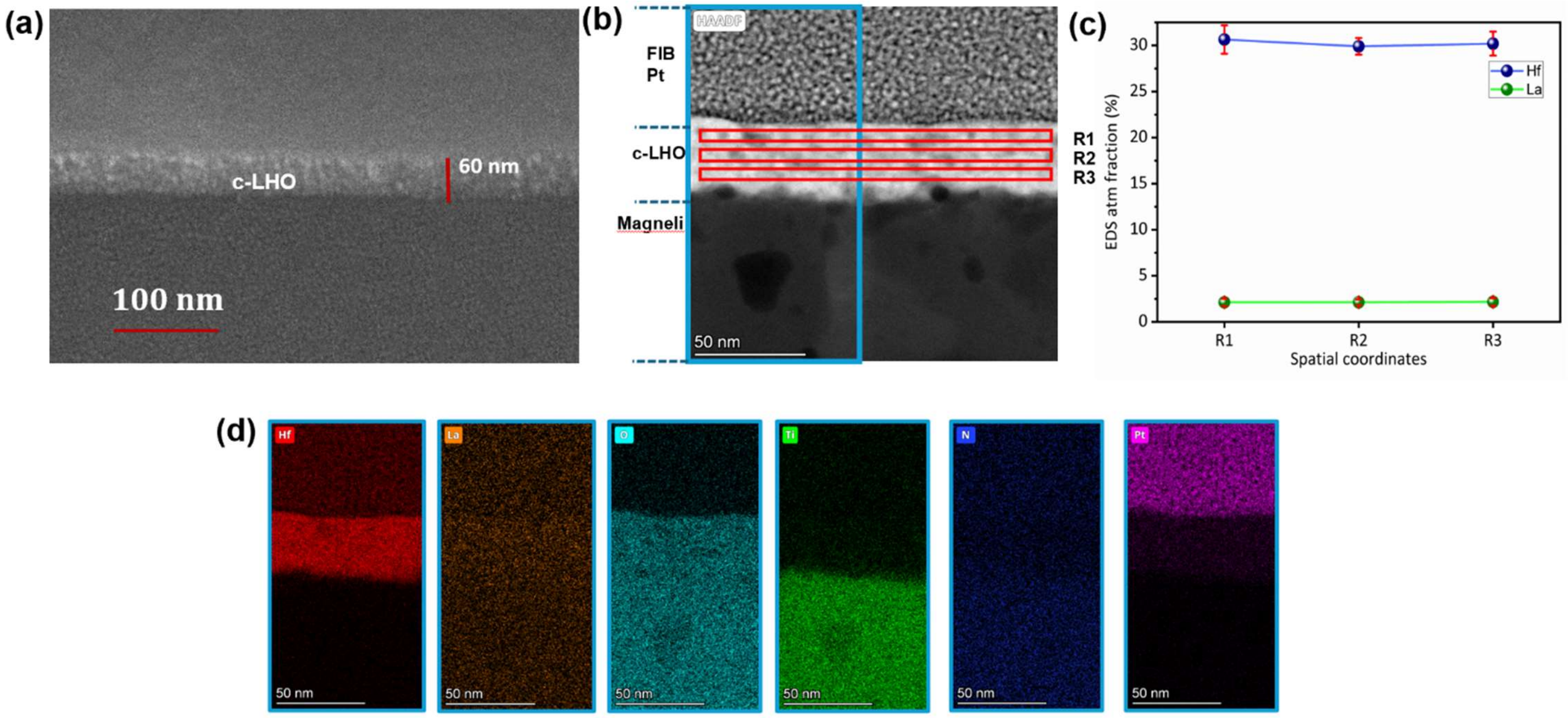

>

**Fig. S2. Chemical uniformity of the crystalline LHO and the magneli layers**

Fig **(a)** FESEM cross-sectional view **(b)** Low magnification HAADF-STEM cross-sectional image showing LHO and the magneli layers. **(b)** Hf and La composition extracted from the EDS spectra obtained from R1, R2 and R3 marked in Fig **(c)**. Note that La:Hf is ~2.1:30 which is about 7% of La in the cationic sites **(d)** Elemental maps distribution of various elements in the stack representing a uniform distribution of Hf and La throughout the LHO region, and Ti and oxygen in the magneli layer. Note that N signal from the magneli layer is very weak.

<

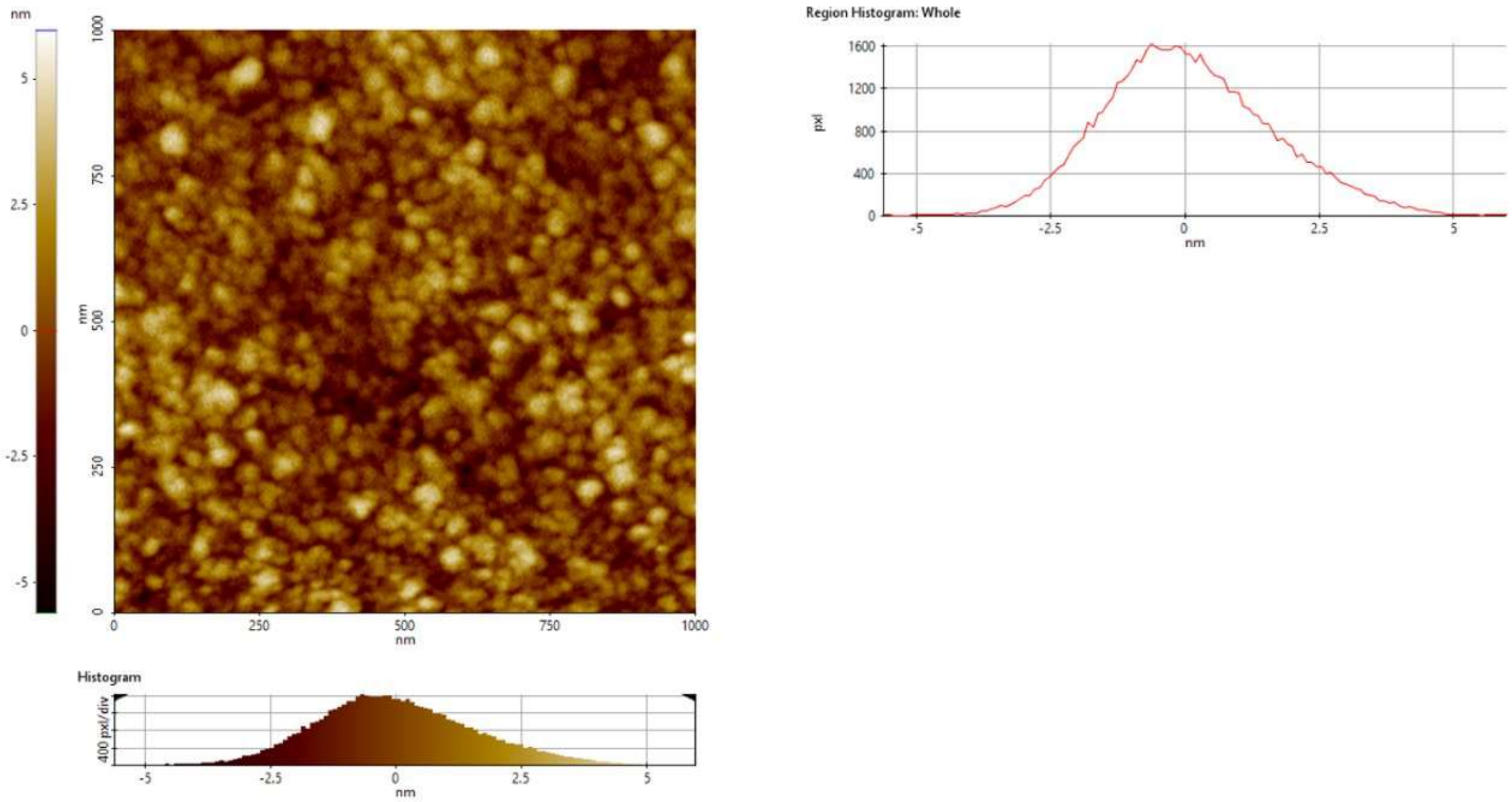

Statistics

| Region | Min(nm) | Max(nm) | Mid(nm) | Mean(nm) | Rpv(nm) | Rq(nm) | Ra(nm) | Rz(nm) | Rsk | Rku |
|---|---|---|---|---|---|---|---|---|---|---|
| Whole | -5.611 | 5.961 | 0.175 | 0.000 | 11.573 | 1.529 | 1.224 | 10.648 | -0.277 | 2.894 |

>

**Fig. S3. Film roughness.**

AFM images of c-LHO film showing the uniform distribution of spherical granular grains and the root mean square (RMS) roughness value of 1.52 nm presenting the quality of the film. No cracks or pinholes are observed in the film.

<

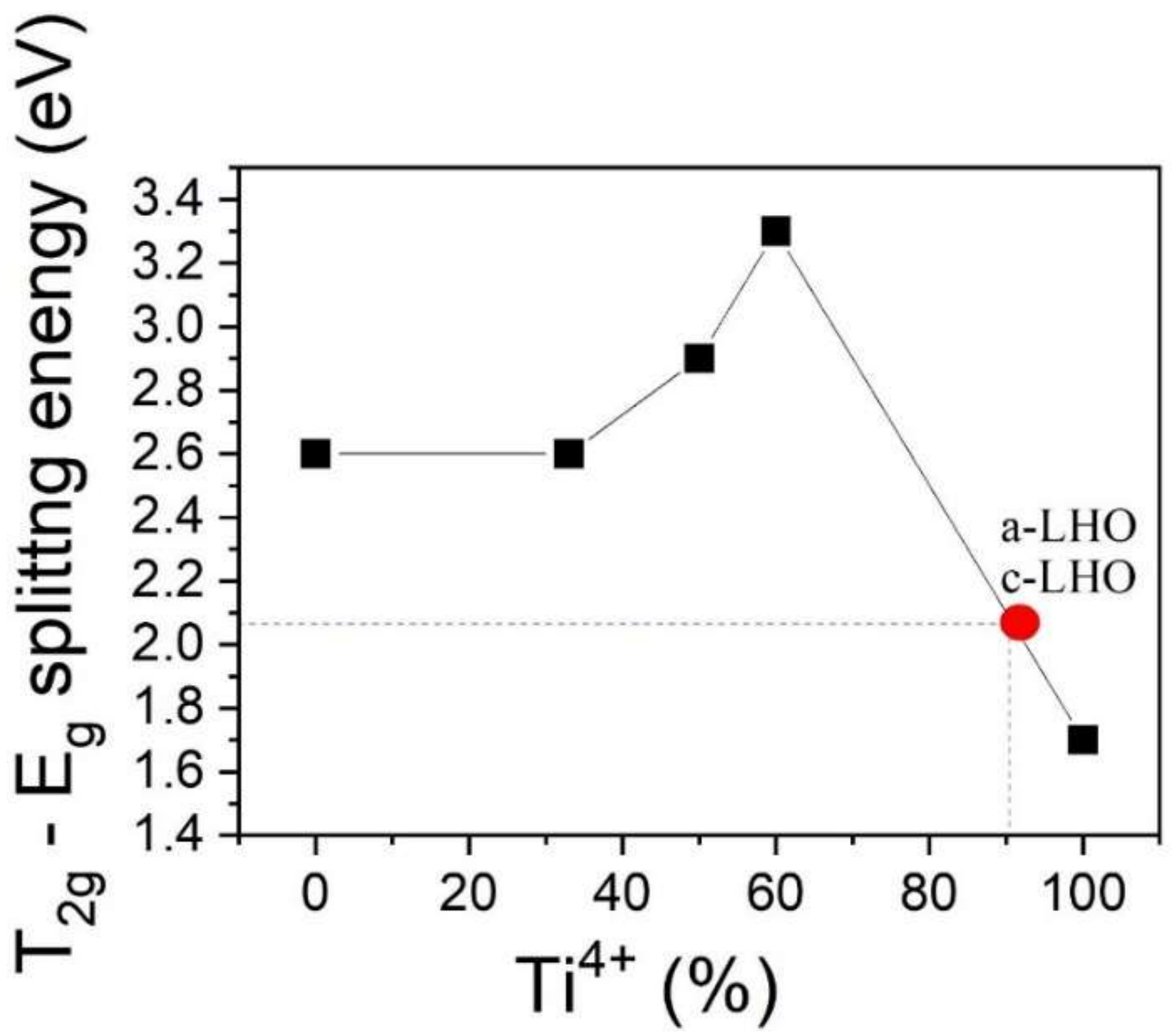

>

**Fig. S4. Determination of $Ti^{3+}$ to Ti $^{4+}$ ratio from XAS, Ti L-edges.**

Energy splitting between $t_{2g}$ and $e_g$ as a function of $Ti^{4+}$ content taken from ref *(35)*. Red circle represents the position of c-LHO/magneli thin film having 90 % Ti $^{4+}$ and 10% Ti $^{3+}$.

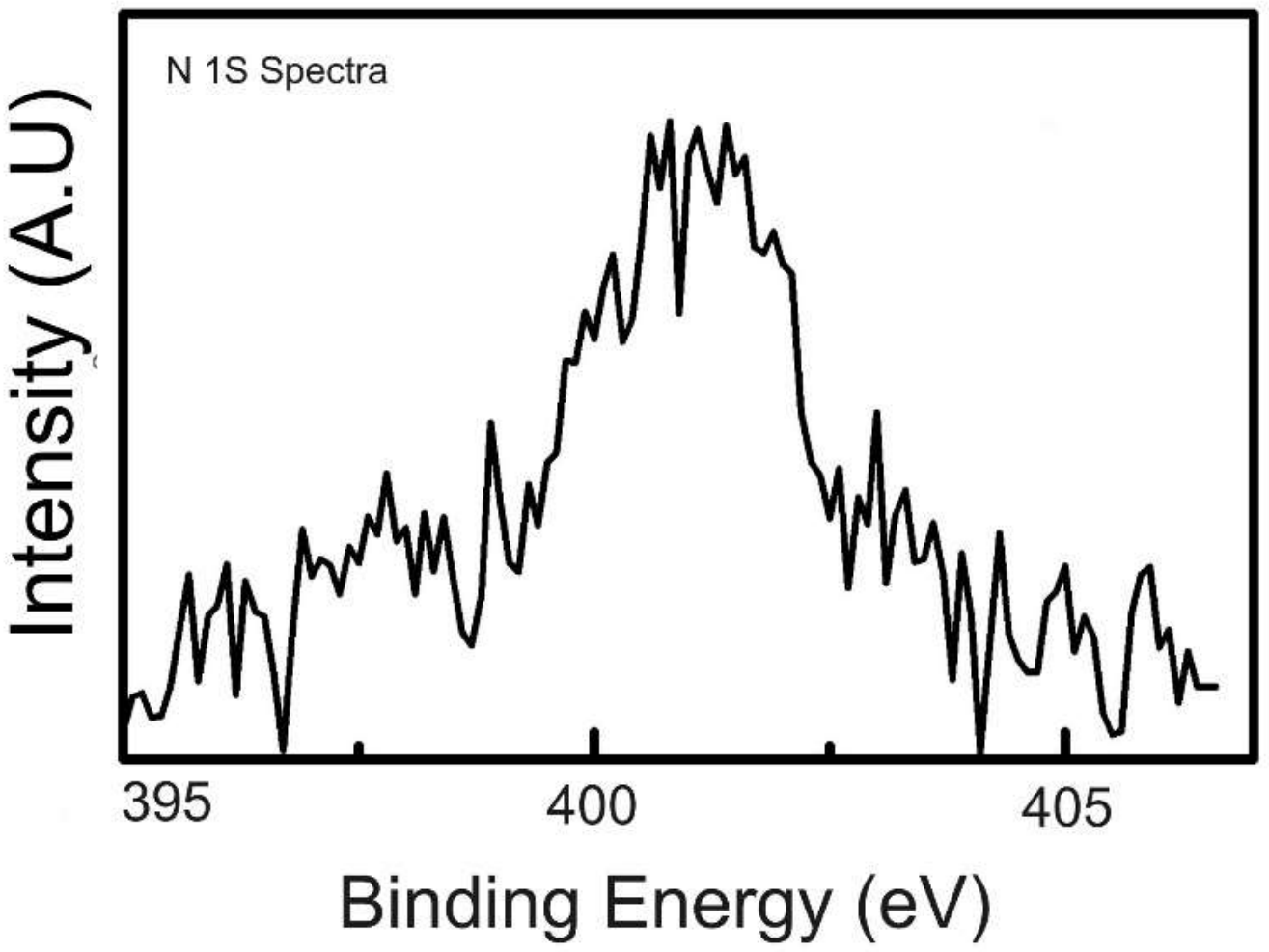

>

**Fig. S5. Trace N detection from XPS**

Depth- resolved XPS spectra of the BE of c-LHO/magneli//Si showing very weak N 1S edge.

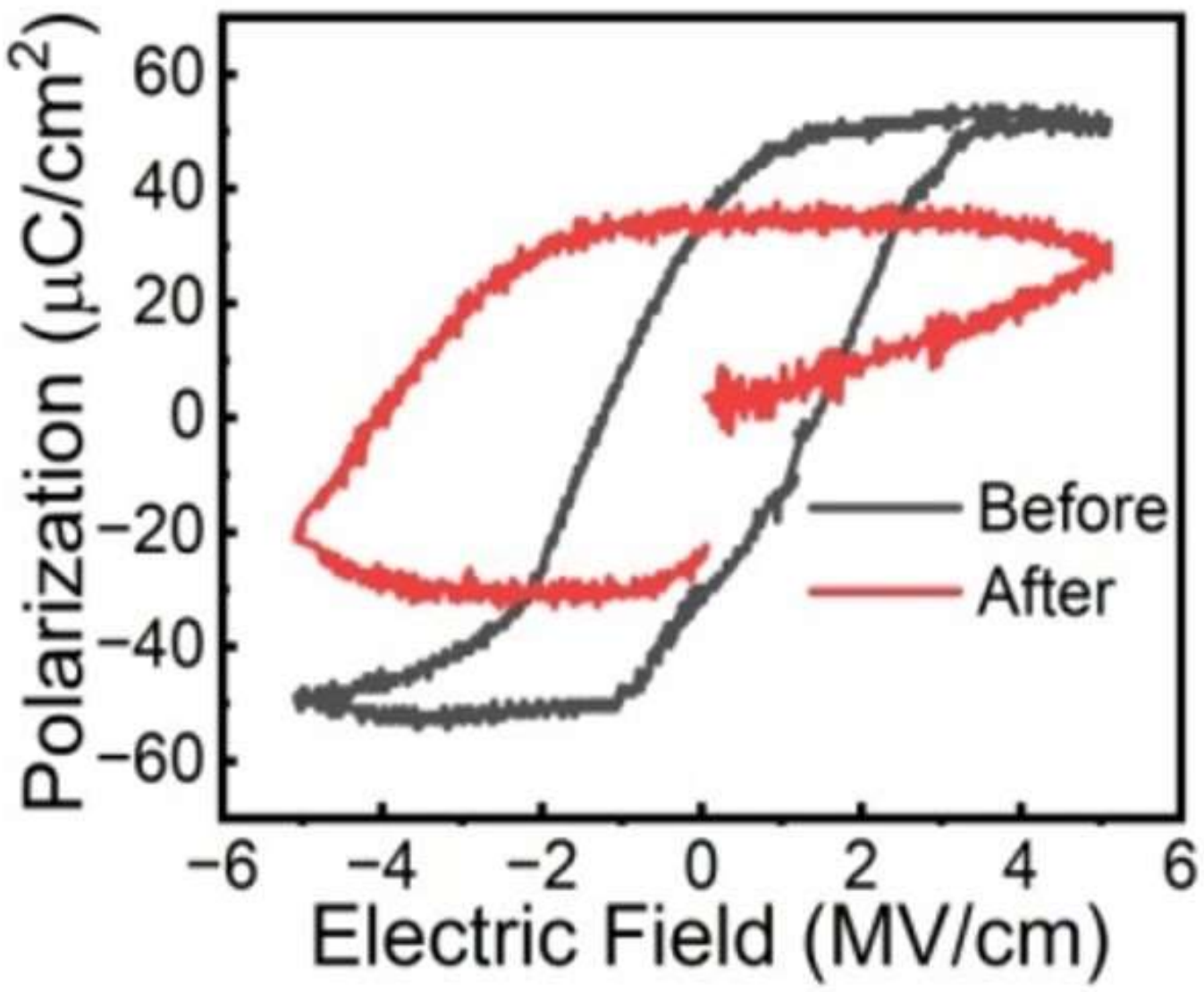

>

**Fig. S6. Ferroelectric hysteresis evolution with fatigue**

P-E hysteresis in the TiN/LHO/magneli layer before cycling for the 1st time (black) and after $10^9$ cycles (red). Some leakage sets in and the saturation polarization goes down upon $10^9$ times cycling.

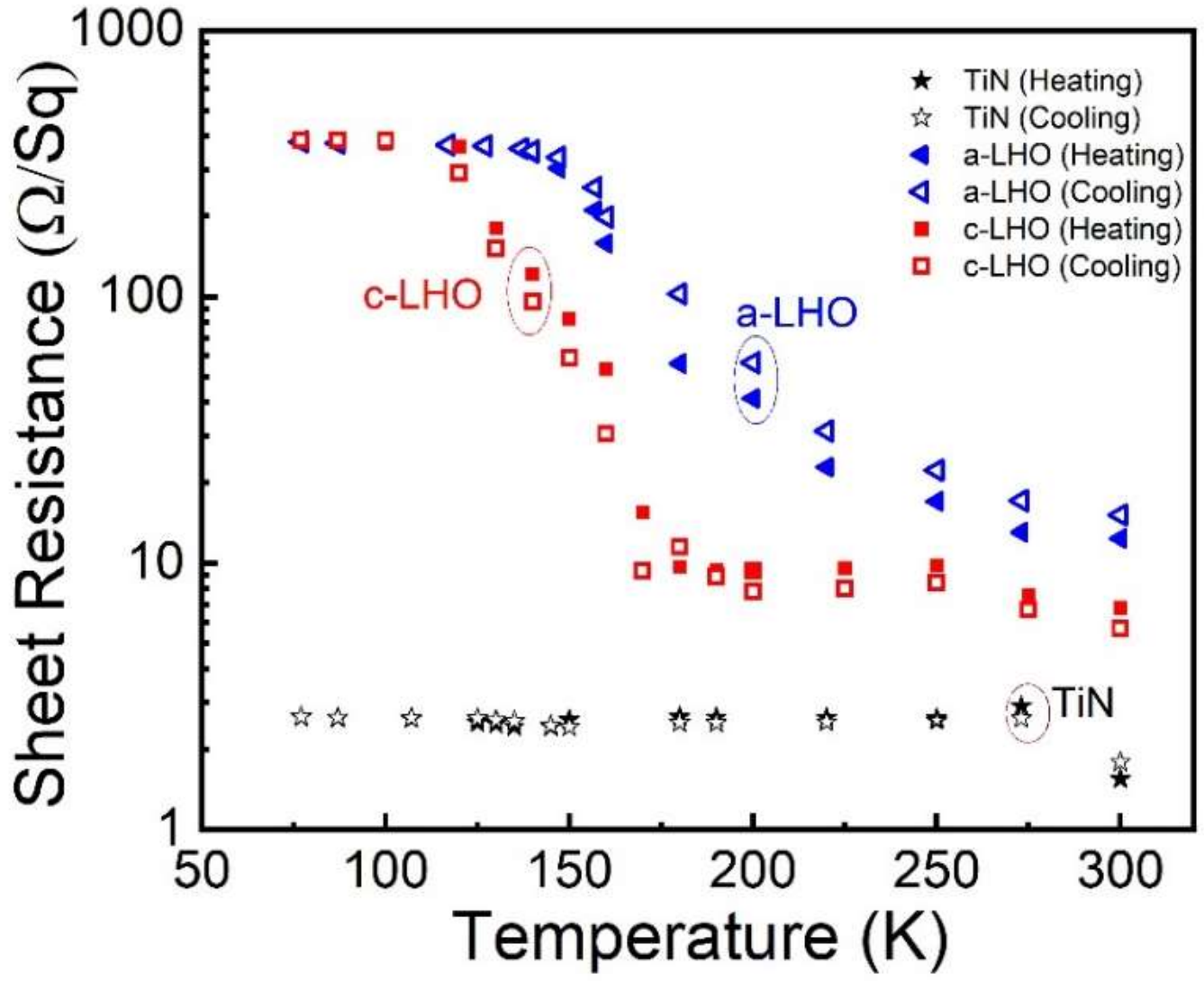

>

**Fig. S7. Reversible metal-insulator transition in the magneli layer**

Temperature-dependent sheet resistance of the reference TiN (black) compared with the sheet resistance of magneli layers interfaced with c-LHO (red) and a-LHO (blue). Note the very small hysteresis, and a slight increase in the transition temperature of the magneli layer when interfaced with a-LHO in comparison with c-LHO. This indicates tunability of $T_C$ of the magneli layer possibly due to slight fluctuations in composition (N content).

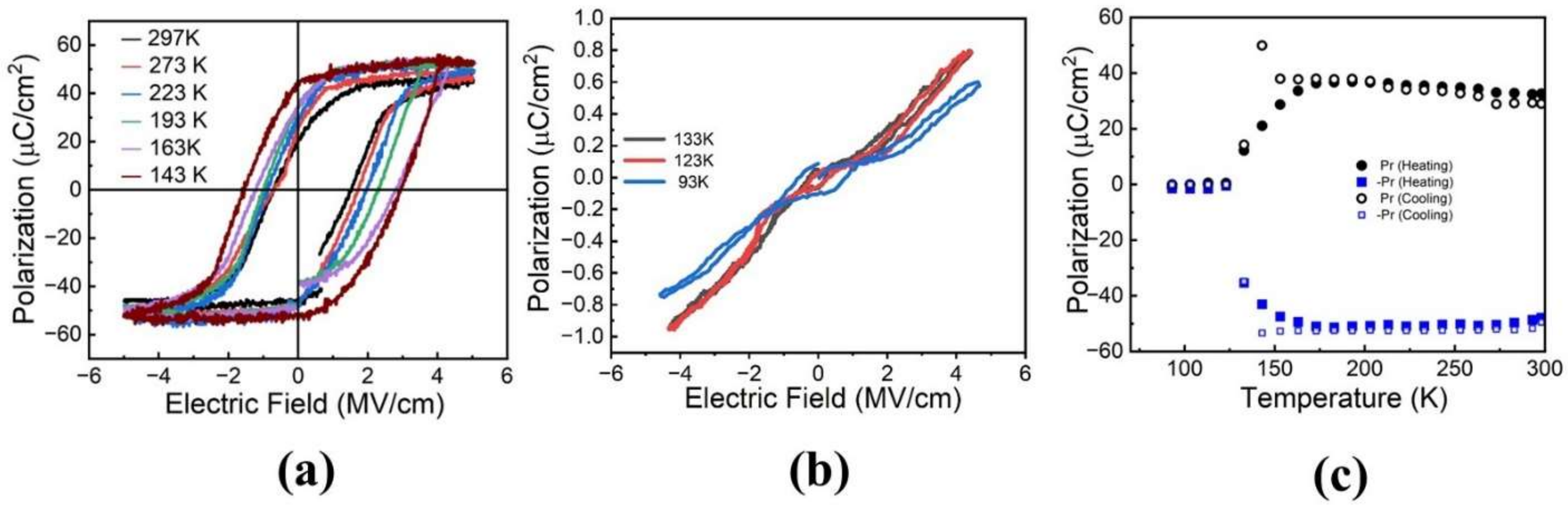

**Fig. S8. Reversible change in polarization in LHO layer, concomitant with the IMT in the magneli layer**

Temperature-dependent P-E hysteresis measured at 3kHz (a) in the temperature range 297 K-143 K and (b) 133 K-93 K. The remnant and spontaneous polarization values decrease as the temperature increases from 143 K-297 K. Hysteresis collapses below 140K and ferroelectricity disappears. (c) Variation of remanent polarization with temperature during heating and cooling cycles. Note the slim hysteresis again, consistent with the slim hysteresis of the sheet resistance of the magneli layer with temperature (Fig S7).

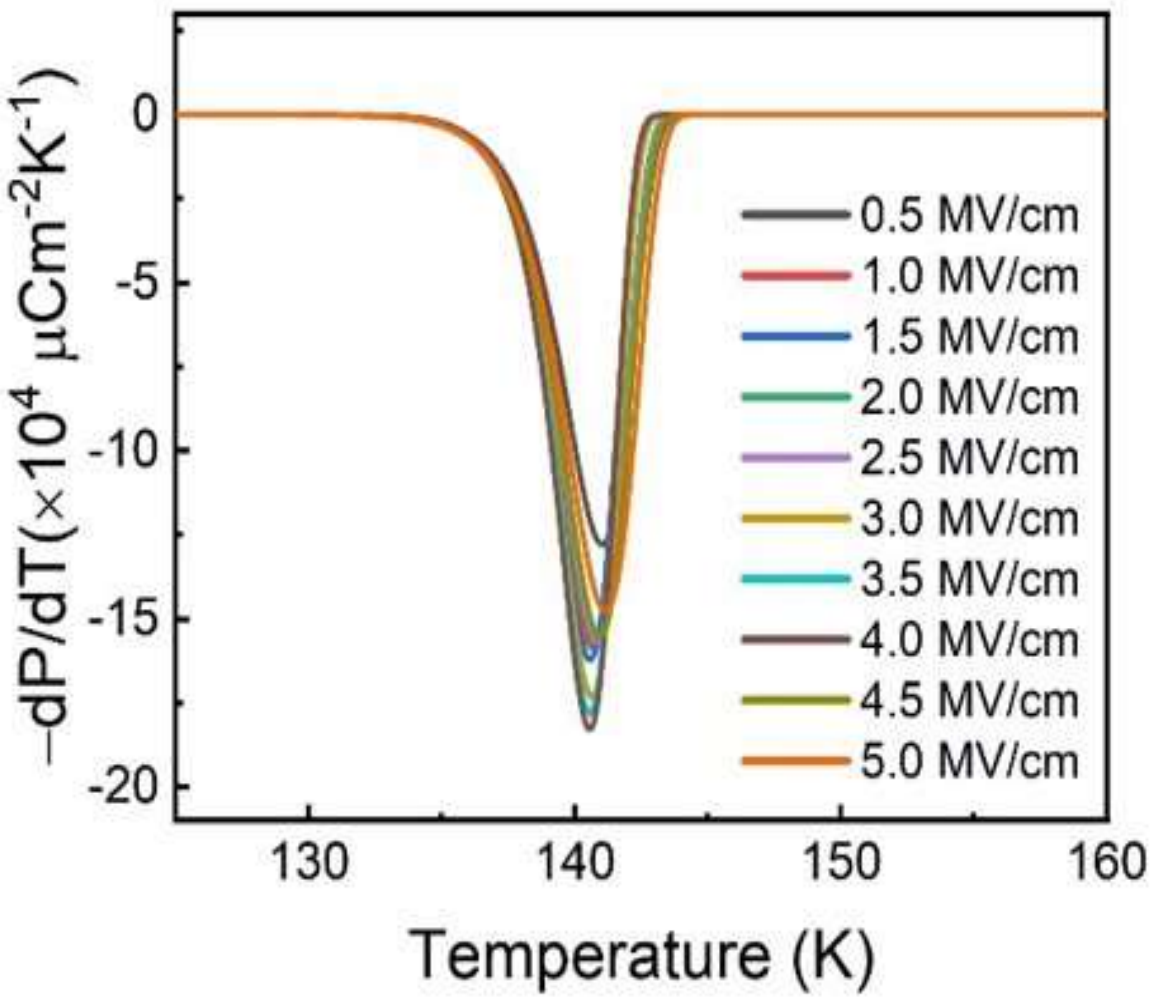

**(a)**

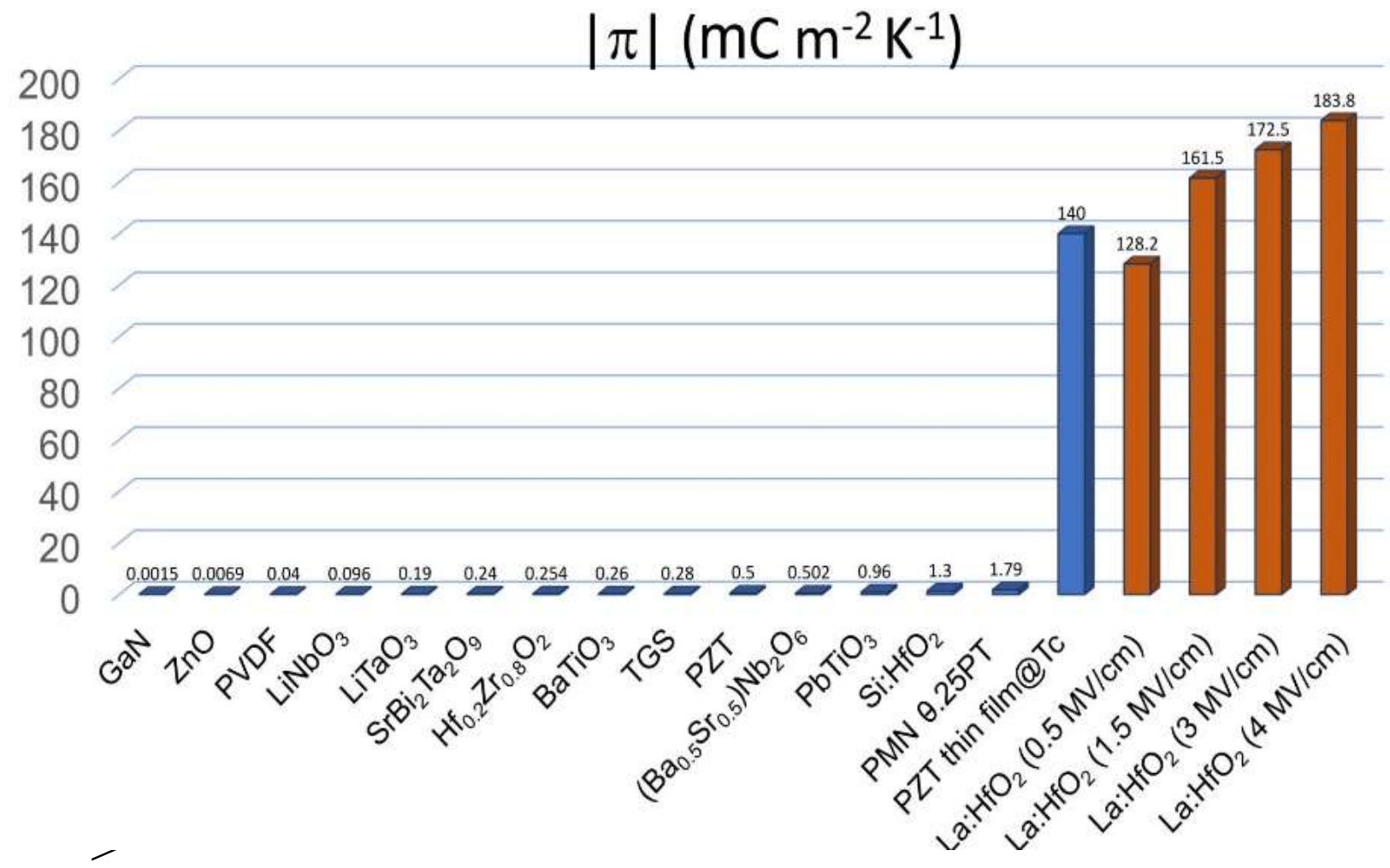

**(b)**

**Fig. S9. Negative pyroelectric coefficient**

**(a)** Variation of pyroelectric coefficient (-dP/dT) as a function of temperature and electric field. The pyroelectric coefficient values are huge at the transition temperature 140 K and at 5 MV $cm^{-1}$. See Fig. S9**(b)** for a bar chart comparing pyroelectric coefficients of various materials *(15-17, 40-49)*.

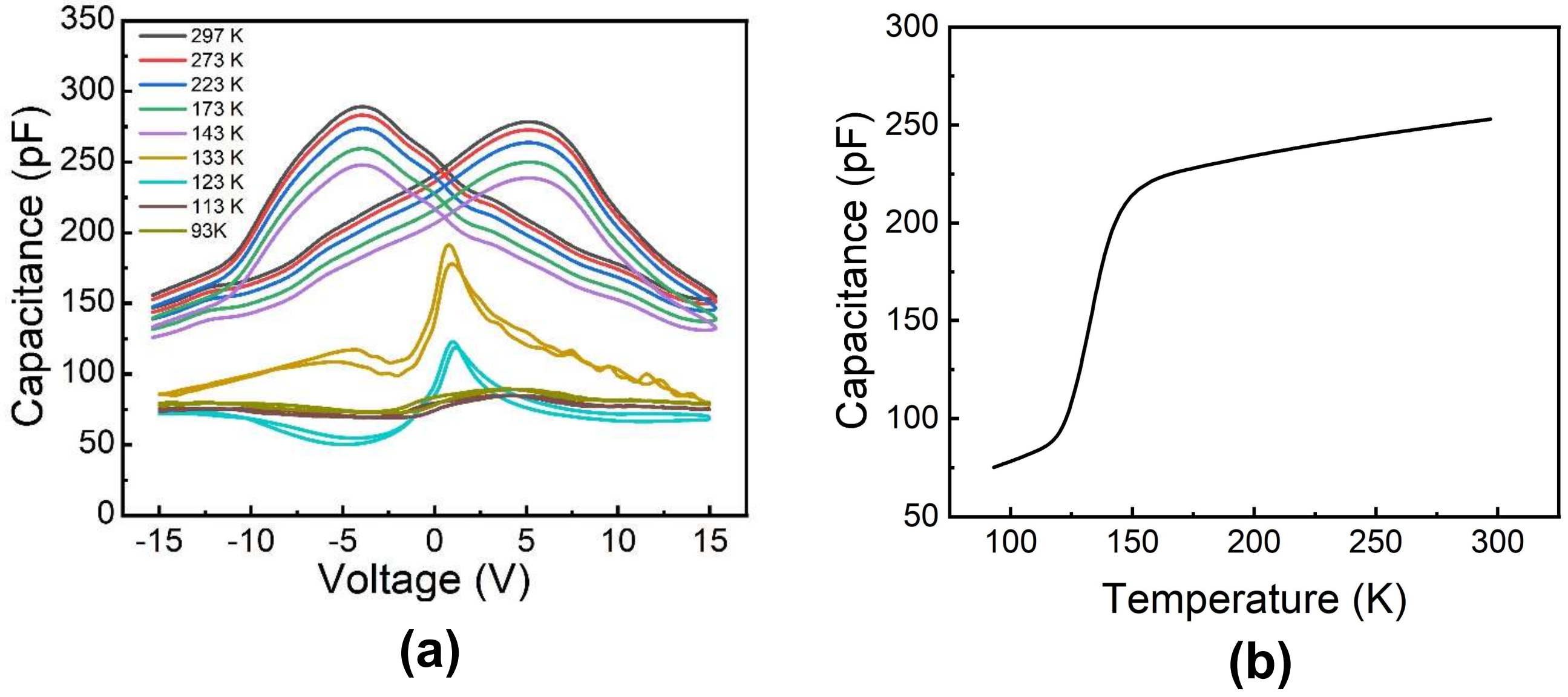

<

**Fig. S10. Evolution of the C-V hysteresis loops as a function of temperature**

**(a)**Temperature-dependent C-V plots of TiN/LHO/magneli capacitors showing ferroelectric butterfly hysteresis loops beyond 143 K, and not below 133 K **(b)** Zero field capacitance plotted as a function of temperature.

>

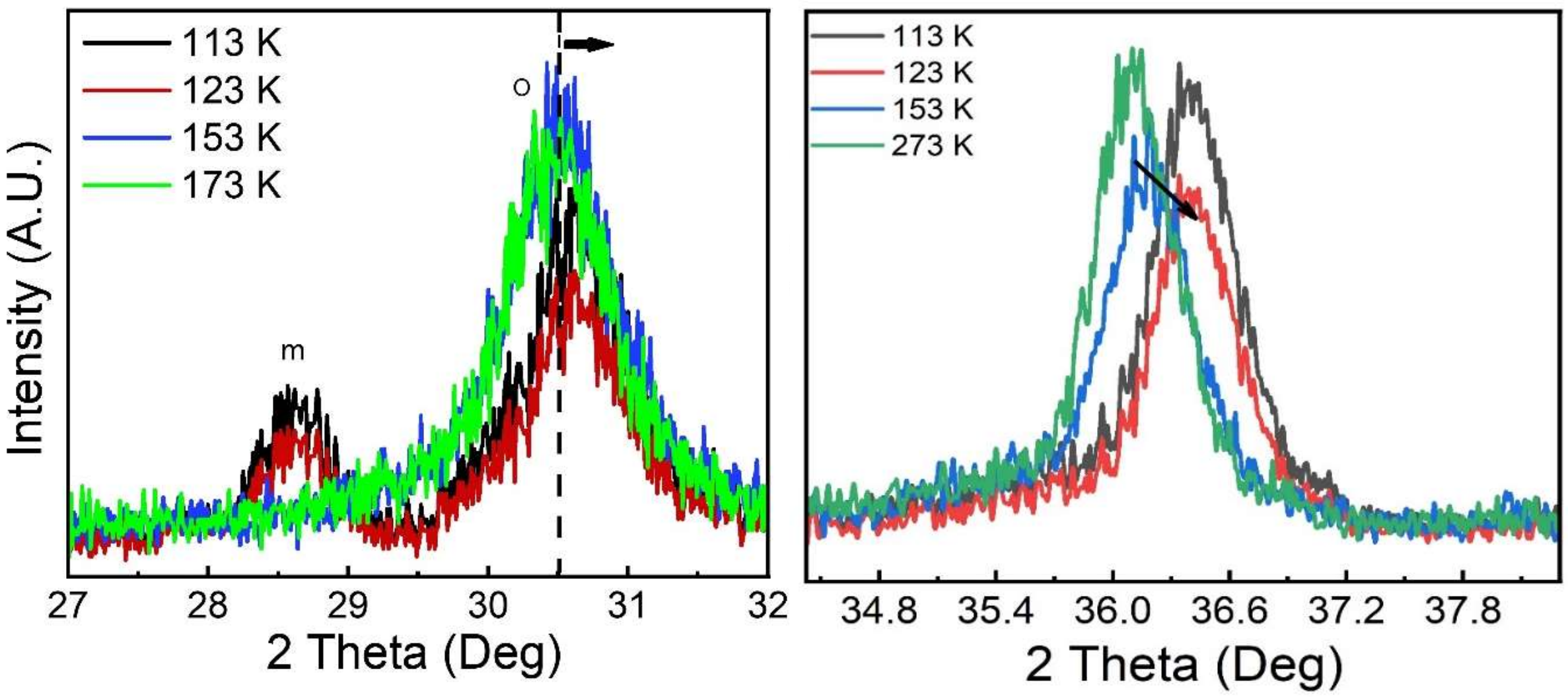

**(a)** **(b)**

**Fig. S11. Evolution of the structure of c-LHO and the magneli layers with temperature**

(a)Temperature-dependent GIXRD patterns with an enlarged view from (a) 27 $^o$-32 $^o$ and (b) 34 $^o$-38 $^o$ representing the emergence of non-polar monoclinic phase at and below 145 K (2θ=28.72 $^o$). The o-(111) peak at 30.42 $^o$ shifts to 30.62 $^o$ representing the polar to nonpolar orthorhombic phase at and below the transition temperature. (b) Enlarged view of XRD patterns from 34$^o$-38$^o$ capturing the magneli phase (120) reflection. A clear shift in the peak occurs from 36.2 $^o$ to 36.4$^o$ at the phase transition (upon cooling from 153 to 123 K).

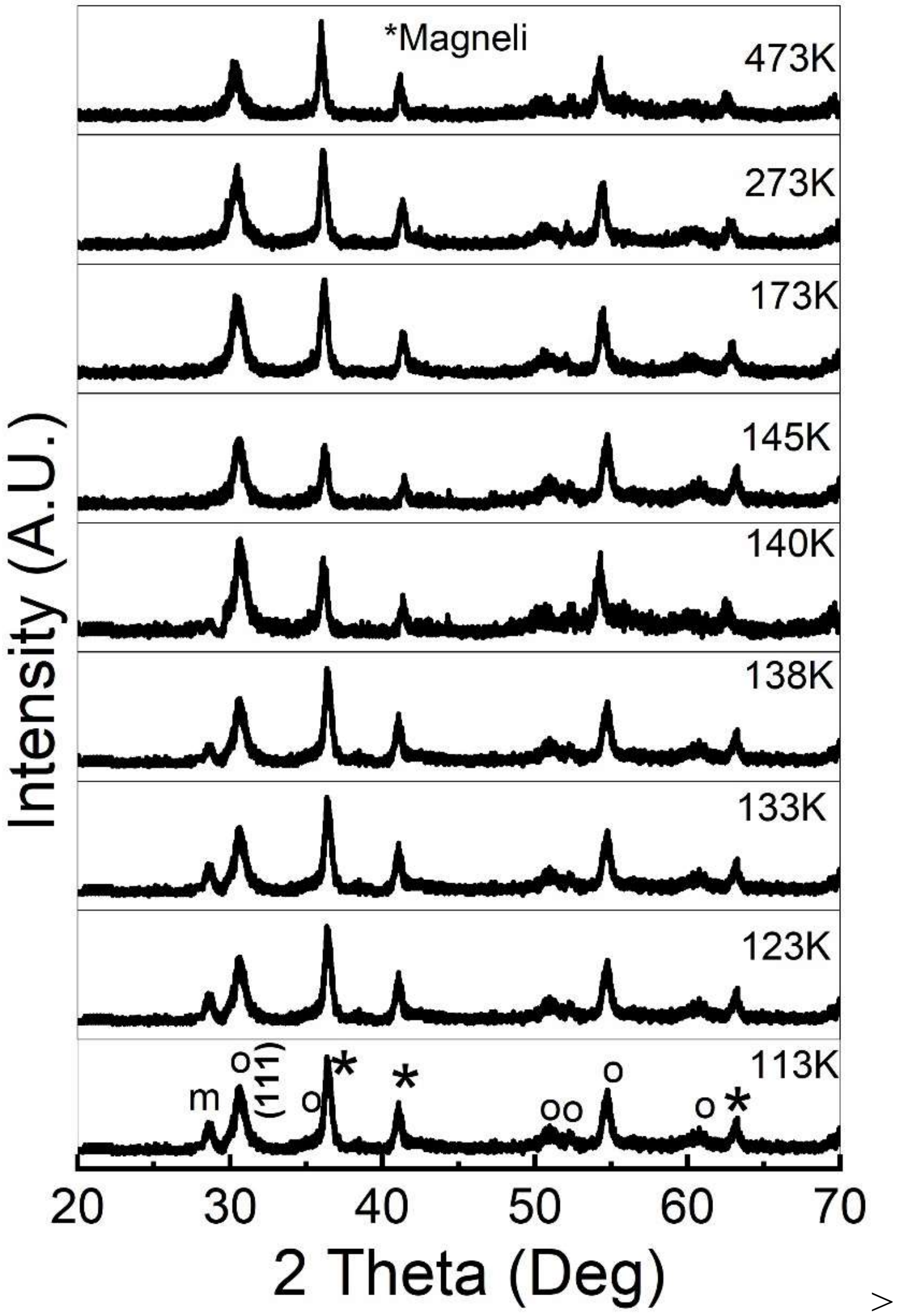

>

**Fig. S12. Evolution of the GIXRD patterns over a larger angle range (20-70 º) as a function of temperature.**

Temperature-dependent GIXRD patterns at varying temperatures representing structural change at 140 K, a combined graph.

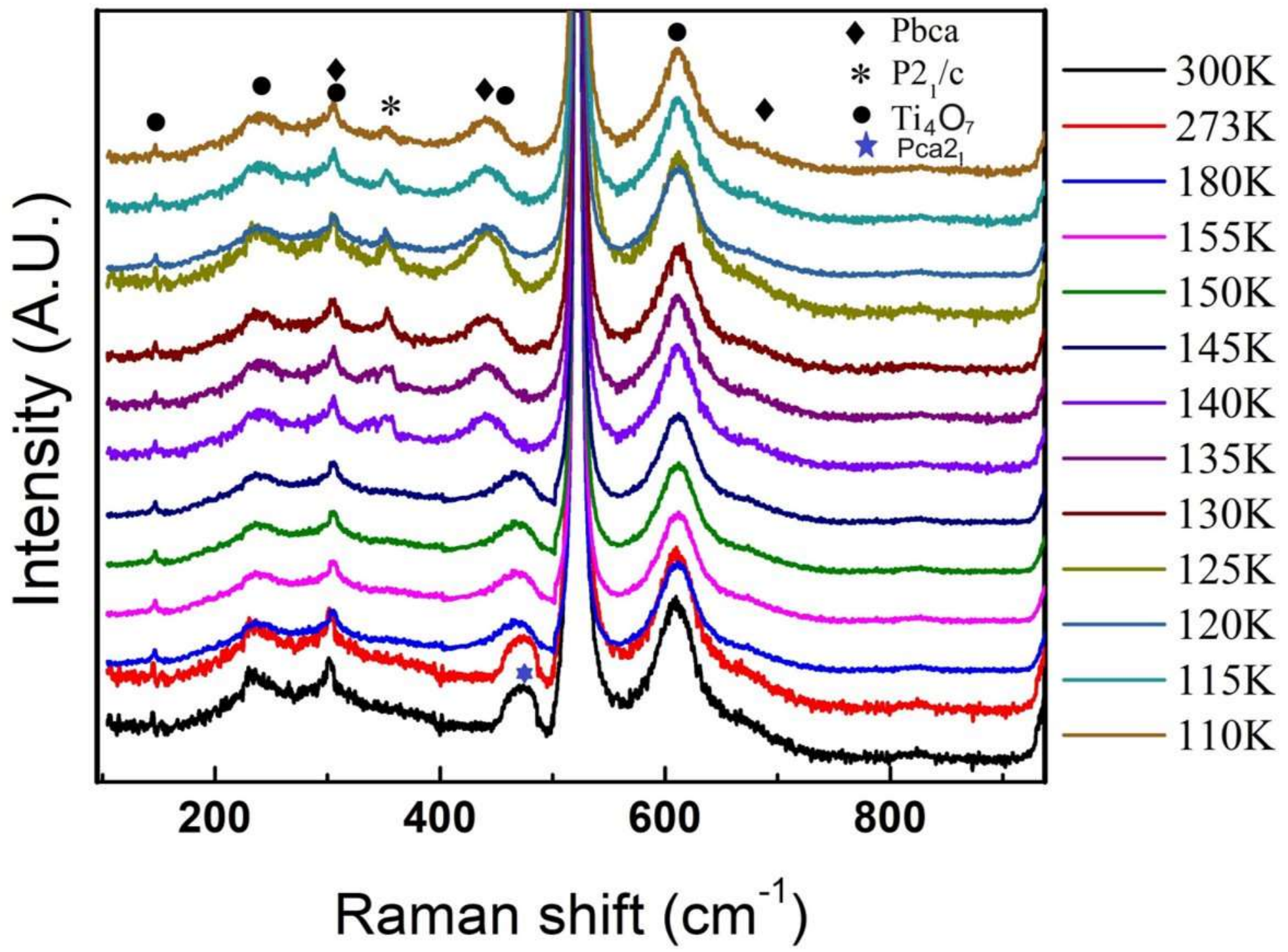

>

**Fig. S13. Temperature-dependent Raman data**

Temperature-dependent Raman spectra for the c-LHO/Magneli stack

>

**Table S1.**

| Raman Frequency (300K) | Phases | Raman Frequency (110K) | Phases |
|---|---|---|---|
| 144.4 | $Ti_4O_7$ | 146.9 | $Ti_4O_7$ |
| 236.3 | $Ti_4O_7$ | 239.19 | $Ti_4O_7$ |
| 301.46 | Si | 304.4 | Si |
| 301.44 | $HfO_2$ ($A_2$) | 304.4 | $HfO_2$ ($B_{1g}$) (Pbca) |
| | | 352.2 | $HfO_2$ ($B_g$) ($P2_1/c$) |
| 362.0 | $HfO_2$ ($A_1$) | | |
| | | 388.0 | $Ti_4O_7$, $HfO_2$ ($A_g$) (Pbca) |
| 393.0 | $HfO_2$ ($A_1$, $B_1$) | | |
| | | 433.7 | $HfO_2$ ($A_g$)(Pbca) |
| 464.4 | $Ti_4O_7$ | 449.9 | $Ti_4O_7$ |
| 476.3 | $HfO_2$ ($A_1$) ($Pca2_1$) | | |
| 520.52 | Si | 523.0 | Si |
| 609.15 | $Ti_4O_7$ | 611.5 | $Ti_4O_7$ |
| 671.8 | $HfO_2$ ($B_2$) | 675.2 | $HfO_2$ ($B_{3g}$) (Pbca) |
| 824.9 | $HfO_2$ ($B_2$) | 827.3 | $HfO_2$ ($B_{3g}$) (Pbca) |

Raman vibrational fingerprints in $HfO_2$ and $Ti_4O_7$ at room temperature and 110 K *(24-27,36,37)*

**Table S2.**

| SAMPLE | T(K) | ΔT(K) | ΔS(J kg$^{-1}$K$^{-1}$) | R ( J kg$^{-1}$) |
|---|---|---|---|---|
| P(VDF-TrFE-CFE) non stretched | 300 | 3.8 | 25 | 3993 |
| P(VDF-TrFE-CFE) non stretched | 300 | 7 | 30 | 5704 |
| P(VDF-TrFE-CFE) Stretched | 300 | 8 | 30 | 4345 |
| P(VDF-TrFE-CFE)/BST67 | 311 | 9.2 | 80 | 10530 |
| P(VDF-TrFE-CFE)/BST71 | 322 | 9.4 | 79 | 10000 |
| P(VDF-TrFE-CFE)/BST74 | 331 | 9.7 | 78.5 | 4065 |
| P(VDF-TrFE-CFE)/BST77 | 337 | 9.9 | 78 | 4180 |
| PST | 341 | 6.2 | 6.3 | 693 |
| 0.93PMN-0.07PT | 298 | 9 | 10 | 135 |
| 0.93PMN-0.07PT | 298 | 13 | 14 | 162 |
| PLZT 11/85/15 | 111 | 12 | 10 | 120 |
| 0.68BFO–0.32BT bulk ceramics | 113 | -2 | -1.9 | - |
| 0.99BT-0.02BMT ceramic* | 416 | 1.21 | 1.22 | - |
| BTO Multilayer Thick Film | 333 | 7.2 | 10.7 | 50 |
| $Pb(Sc_{1/2}Ta_{1/2})O_3$ | / | 2.1-2.3 | 3.44 | - |
| $PbZr_{0.95}Ti_{0.05}O_3$ | 495 | 12 | 8 | 96 |
| $Pb_{0.8}Ba_{0.2}ZrO_3$ | 290 | 45.3 | 46.9 | 2125 |
| $Hf_{0.2}Zr_{0.8}O_2$ | 298 | 13.4 | 16.7 | - |
| P(VDF-TrFE)55/45 | 353 | 12.6 | 60 | 756 |
| $PbSc_{0.5}Ta_{0.5}O_3$ | 341 | 6.2 | 6.3 | 39 |
| PMN-PT 90/10 | 348 | 5 | 5.6 | 28 |
| $Pb(Mg_{1/3}Nb_{2/3})_{0.65}Ti_{0.35}O_3$ | 413 | 31 | 32 | 992 |
| $Ba(Zr_xTi_{1-x})O_3$ | 343 | -4.2 | 7.3 | 30.66 |
| $PbZr_{0.53}Ti_{0.47}O_3/CoFe_2O_4$ | 182 | -52.2 | -94.23 | - |
| PST | 341 | -5 | -5.3 | 695 |
| PLZT | 318 | 40 | 50 | 2000 |
| P(VDF-TrFE) 68/32 | 306 | 20 | 95 | 1900 |
| 5.6 mol% Si doped $HfO_2$ | 298 | 9.5 | 8.9 | 203 |
| $Hf_{0.5}Zr0.5O_2$ | 448 | -10.8 | -10.9 | - |
| 5.7% Al-doped $HfO_2$ | 80 | -7.4 | -7.8 | - |
| Pt/BTO/SRO FTJs | 300 | 4.5 | - | ~45 |
| $SrTiO_3$ | 242 | - | 0.35 | |
| **TiN/La:$HfO_2$/magneli//Si (PRESENT STUDY)** | 140 | 106 | 8000 | 25000 |

Electrocaloric metric parameters for different materials systems *(7,8,18, 38,50-53)*.

**Acknowledgement**

**Funding:** The major part of this work was carried out at Micro and Nano Characterization Facility (MNCF), and National Nanofabrication Center (NNFC) located at CeNSE, IISc Bengaluru, and benefitted from all the help and support from the staff and all authors acknowledge the usage of NNFC, MNCF, and Advanced Facility for Microscopy and Microanalysis of IISc for various fabrication and characterization studies. P.N. acknowledges Start-up grant from IISc, Infosys Young Researcher award, and SERB (DST), New Delhi, Govt. of India CRG/2022/003506. JMA acknowledges IISC-IOE Postdoctoral Fellowship and IEEE-ISAF-ISIF-PFM travel grant 2023. PN and JMA acknowledge Prof. Pierre-Eymeric Janolin, Université Paris-Saclay, CNRS, Centrale Supélec, Laboratoire SPMS and Prof. Brahim Dkhil, Centrale Supélec, Université Paris-Saclay for their constant support and fruitful discussions. BKD acknowledge Dr. Mukul Gupta and Dr. Uday Deshpande for XAS and XPS measurements and discussions.

**Author contributions:**
P.N. and J.M.A, conceived the idea. J.M.A synthesized the samples and made the MIM devices. J.M.A performed the XRD, FESEM, AFM, Raman, electric, dielectric and ferroelectric testing on the samples and J.M.A and P.N. analyzed the data. S.K.P carried out TEM imaging and S.K.P., P.N., and J.M.A. performed the data analysis. B.D. carried out XAS experimentation and analysis. J.M.A and SDK calculated the pyroelectric and electrocaloric parameters from the temperature-dependent polarization data. S.D.K. performed the data fitting to calculate electrocaloric parameters. All the authors discussed the data. J.M.A. and P.N. co-wrote the manuscript, which was reviewed, edited, and approved by all the authors.

**Competing interests:** The authors declare that they have no competing interests.

**Data and materials availability:** Data is available on request from the corresponding authors.